\documentclass[aps,pra, twocolumn, superscriptaddress, longbibliography]{revtex4-2}
\usepackage{times}
\usepackage{amsmath}
\usepackage{amssymb}
\usepackage{graphicx}
\usepackage[caption=false, position=top, singlelinecheck=off, justification=raggedright]{subfig}
\usepackage{color}
\usepackage{pst-node}
\usepackage[unicode]{hyperref}
\hypersetup{unicode=true, colorlinks=true, linkcolor=blue, citecolor=blue}
\usepackage{float}
\usepackage[normalem]{ulem}

\begin{document}

\title{Realizing limit cycles in dissipative bosonic systems}

\author{Jim Skulte}
\affiliation{Center for Optical Quantum Technologies and Institute for Quantum Physics, Universit\"at Hamburg, 22761 Hamburg, Germany}
\affiliation{The Hamburg Center for Ultrafast Imaging, Luruper Chaussee 149, 22761 Hamburg, Germany}

\author{Phatthamon Kongkhambut}
\affiliation{Center for Optical Quantum Technologies and Institute for Quantum Physics, Universit\"at Hamburg, 22761 Hamburg, Germany}

\author{Hans Ke{\ss}ler}
\affiliation{Center for Optical Quantum Technologies and Institute for Quantum Physics, Universit\"at Hamburg, 22761 Hamburg, Germany}
\affiliation{Physikalisches Institut, Rheinische Friedrich-Wilhelms-Universität, 53115 Bonn, Germany}

\author{Andreas Hemmerich}
\affiliation{Center for Optical Quantum Technologies and Institute for Quantum Physics, Universit\"at Hamburg, 22761 Hamburg, Germany}
\affiliation{The Hamburg Center for Ultrafast Imaging, Luruper Chaussee 149, 22761 Hamburg, Germany}

\author{Ludwig Mathey}
\affiliation{Center for Optical Quantum Technologies and Institute for Quantum Physics, Universit\"at Hamburg, 22761 Hamburg, Germany}
\affiliation{The Hamburg Center for Ultrafast Imaging, Luruper Chaussee 149, 22761 Hamburg, Germany}

\author{Jayson G. Cosme}
\affiliation{National Institute of Physics, University of the Philippines, Diliman, Quezon City 1101, Philippines}

\date{\today}

\begin{abstract}   
We propose a general mechanism for generating limit cycle (LC) oscillations by coupling a linear bosonic mode to a dissipative nonlinear bosonic mode. By analyzing the stability matrix, we show that LCs arise due to a supercritical Hopf bifurcation. We find that the existence of LCs is independent of the sign of the effective nonlinear interaction. The bosonic model can be realised in three-level systems interacting with a quantised light mode as realised in atom-cavity systems. Using such a platform, we experimentally observe LCs for the first time in an atom-cavity system with attractive optical pump lattice, thereby confirming our theoretical predictions for the minimal model and interactions needed to generate LCs for a class of driven-dissipative systems.
\end{abstract}

\maketitle 

\section{Introduction}

A central focus of quantum optics is the understanding of few-level systems coupled to a single photonic mode \cite{walls2007}. A quintessential example is the Dicke model \cite{Dicke1954,DickeModel}, in which a large number of two-level systems are coupled to the same light mode, giving rise to exciting physical phenomena, such as super- and subradiance and the Dicke phase transition. Furthermore, as was pointed out more recently, if this model is extended to incorporate multi-level systems and dissipation, the resulting many-body dynamics can give rise to LCs \cite{Chan2015, Owen2018, Colella2022, Berislav2022, wu2023, wadenpfuhl2023,weis2023exceptional} and continuous time crystals (CTCs) \cite{Iemini2018, Hans2019, Buca2019, Bakker2022, Krishna2023, Kongkhambut2022, Liu2023,Chen2023}. 
LCs are closed phase space trajectories, inherently robust
against noise or perturbations in the initial state. They emerge via continuous time
translation symmetry breaking, manifesting in an oscillatory motion despite the
absence of an explicit time-dependence in their equations of motion. A LC phase in a
many-body system with an unbiased distribution of the time phase of its oscillatory
motion is a CTC and has been recently demonstrated experimentally in a
continuously driven atom-cavity system \cite{Kongkhambut2022}.

\begin{figure}[!ht]
\centering
\includegraphics[width=1\columnwidth]{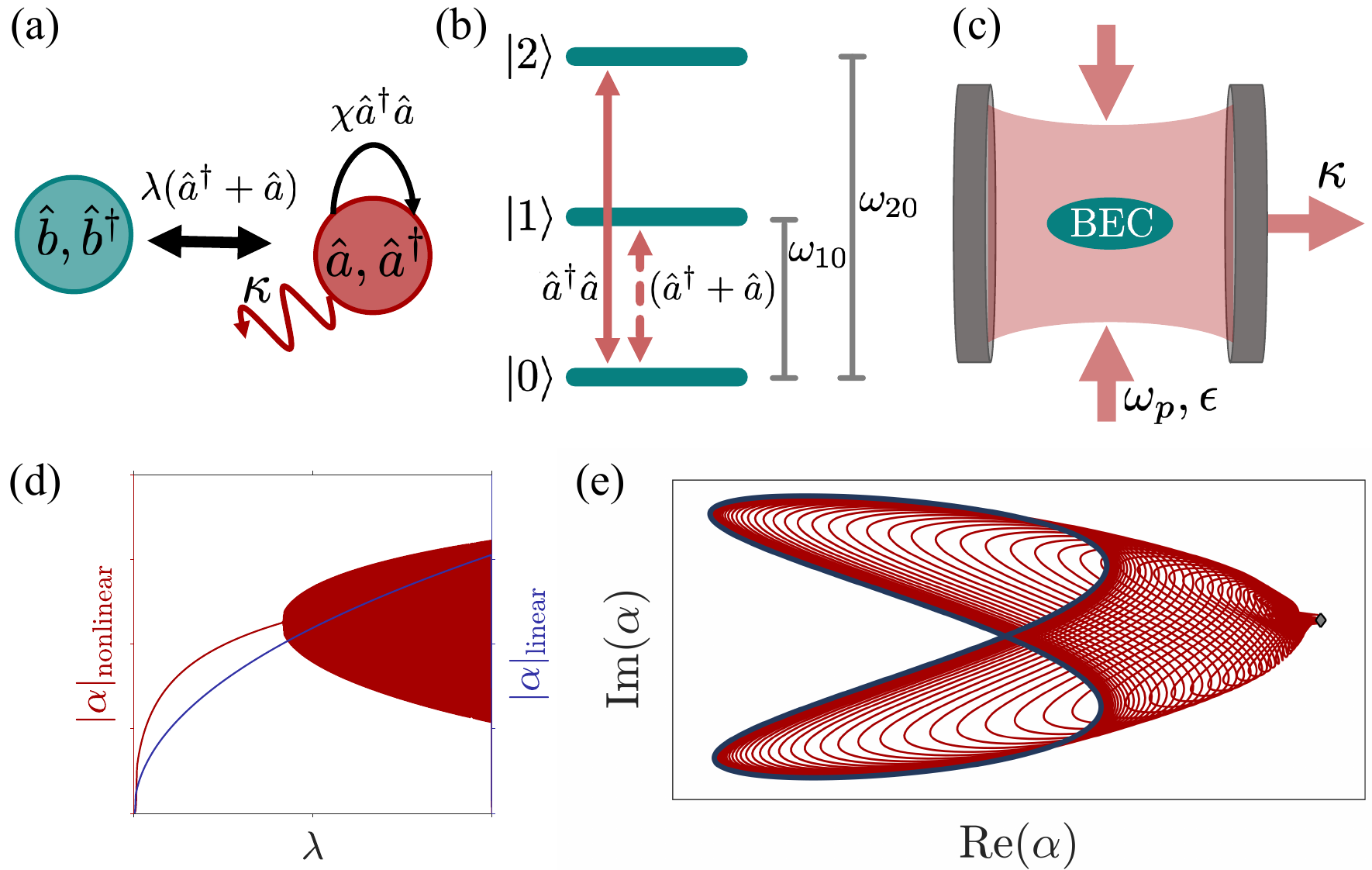}
\caption{(a) A bosonic mode (teal) $\hat{b}$ interacts with a dissipative bosonic mode (red) $\hat{a}$ with a nonlinearity proportional to $\chi$ and damping $\kappa$. The width of the arrows denote the strength of the interactions. (b) Approximate three-level model for (c) a BEC (teal) coupled to a single light mode including single-photon coupling with strength $\lambda$ and Kerr nonlinearity with strength $\chi$. In (c), the BEC is transversely pumped by a standing wave potential formed by two laser beams and placed inside a high-finesse cavity. The rate of emitted light from the cavity is $\kappa$. (d) Exemplary dynamics of the Dicke model and the Dicke model including a Kerr-like nonlinearity for varying coupling strength $\lambda$. While the Dicke model reaches a steady state after entering the superradiant phase, the nonlinear Dicke model enters the limit cycle phase for increasing $\lambda$. (e) Dynamics of the light field in phase space starting from a fixed point (gray diamond) and relaxing towards the limit cycle (blue line).} 
\label{fig:1} 
\end{figure}

In this work, we put forth a model, giving rise to LC oscillations, that can either be
understood in terms of a collection of two-level systems coupled to a non-linear
photonic mode, or a collection of three-level systems, coupled to a linear photonic
mode. The transition between the representations with three- and two-level systems
arises by adiabatically eliminating one of the levels in the three-level systems and
thereby generating a nonlinearity in the photonic mode. The LC behavior is striking
since additional quantum modes beyond the two-level approximation increase the
complexity of a quantum system, and thus are expected to support ergodicity in
generic systems \cite{zaletel2023}. To obtain a concrete implementation of LC dynamics, we map
our generic model onto an atom-cavity system, and show that the current
understanding, that the emergence of LC phases in atom-cavity systems necessarily
relies on the use of a repulsive light-shift potential, is incomplete \cite{Keeling2010, Bhaseen2012, Piazza2015, Kongkhambut2022}. In this
paper, we elucidate that the fundamental mechanism is in fact a Kerr-like nonlinearity
for the photons, which is induced by a third atomic level, typically neglected in
standard Dicke-like models of atom-cavity systems. This nonlinearity can arise
irrespective of the sign of the light-shift potential or pump-atom detuning, such that an
LC phase may emerge also for negative coupling parameters, which we
experimentally demonstrate in this work.

We expect our results to apply to a wide class of systems, wherein the interactions are mediated by a bosonic mode, such as in cavity-magnon systems \cite{Zare2022} and superconducting circuits \cite{Chang2020, Minganti2023}, provided that they satisfy the form of the coupling between the modes in our effective bosonic model as schematically depicted in Figs.~\ref{fig:1}(a) and \ref{fig:1}(b). {In particular, the Kerr-like nonlinearity needed for the limit cycle to emerge could originate from either a density- or intensity-dependent coupling of a degree-of-freedom and the dissipative mode. Some examples include a Kerr medium coupled to a cavity field \cite{larson2021} and photon-phonon coupling in cavity-optomechanical systems \cite{aspelmeyer2014,Rameshti2022}.}

{The paper is organized as follows. In Sec.~\ref{sec:theory}, we discuss the minimal model and use bifurcation theory to explore the instabilities in the system. In Sec.~\ref{sec:expt}, we discuss the atom-cavity implementation and present the experimental results showing the emergence of limit cycles in an atom-cavity system with attractive light-shift pump potential. Finally, we conclude in Sec.~\ref{sec:conc}.}

\section{Theory}\label{sec:theory}
\subsection{General model}

{
We consider a general model describing three bosonic modes ($a$, $b$, and $c$) with the $a$-mode being dissipative as its occupation decays at a rate of $\kappa$. The three-mode Hamiltonian is
\begin{align}
\label{eq:H}
\hat{H} &= \omega_p \hat{a}^\dagger \hat{a} + \omega_{10} \hat{b}^\dagger \hat{b}+\omega_{20} \hat{c}^\dagger \hat{c} +\lambda \left(\hat{a}^\dagger+\hat{a} \right)\left(\hat{b}^\dagger+\hat{b} \right) \\ \nonumber
&\qquad + \chi \hat{a}^\dagger \hat{a} \left(\hat{c}^\dagger+\hat{c} \right).
\end{align}
The natural frequencies of the three modes are $\omega_p$, $\omega_{10}$, and $\omega_{20}$. The $b$-mode interacts with the $a$-mode via an amplitude-dependent coupling with strength $\lambda$. On the other hand, a density- or intensity-dependent interaction characterized by $\chi$ couples the $a$- and $c$-modes.
}

{
Applying mean-field theory by setting $\langle \hat{a} \rangle=\alpha$, $\langle \hat{b} \rangle=\beta$, $\langle \hat{c} \rangle=\gamma$, and $\langle \hat{A}\hat{B}  \rangle \approx \langle \hat{A} \rangle \langle \hat{B}  \rangle$, we obtain the following set of equations of motion (EOM) for the three-mode system
\begin{align}\label{eq:eom3mode}
\frac{\mathrm{d} \alpha}{\mathrm{d} t} &= -i \left[ \omega_p-i\kappa+\chi \left(\gamma+\gamma^* \right) \right] \alpha-i\lambda \left(\beta+\beta^* \right)  \\ \nonumber
\frac{\mathrm{d} \beta}{\mathrm{d} t} &= -i \omega_{10} \beta -i \lambda  \left(\alpha+\alpha^* \right) \\ \nonumber
 \frac{\mathrm{d} \gamma}{\mathrm{d} t} &= -i \omega_{20} \gamma -i \chi  \alpha^* \alpha~.
\end{align}
}

{
We can adiabatically eliminate the $c$-mode for $\omega_{20} \ll \omega_{10},\omega_p$, such that we approximate $\mathrm{d} \gamma / \mathrm{d} t \approx 0$ in the last line of Eq.~\eqref{eq:eom3mode}. This yields an expression for $\gamma$ given by
\begin{equation}
\gamma = -\frac{\chi}{\omega_{20}} |\alpha|^2~.
\end{equation}
Using this in the equation for the dissipative mode in the first line of Eq.~\eqref{eq:eom3mode}, we obtain an effective two-mode EOM
\begin{align}
\frac{\mathrm{d} \alpha}{\mathrm{d} t} &= -i \left[ \omega_p-2 \frac{ \chi^2}{\omega_{20}}|\alpha|^2-i\kappa  \right] \alpha-i\lambda \left(\beta+\beta^* \right) \notag  \\ \frac{\mathrm{d} \beta}{\mathrm{d} t} &= -i \omega_{10} \beta -i \lambda  \left(\alpha+\alpha^* \right),
\label{eq:eomhp}
\end{align}
Quantising the remaining modes, an effective Hamiltonian corresponding to Eq.~\eqref{eq:eomhp} reads
\begin{align}
\hat{H} &= \omega_p \hat{a}^\dagger \hat{a} + \omega_{10} \hat{b}^\dagger \hat{b} +\lambda \left(\hat{a}^\dagger+\hat{a} \right)\left(\hat{b}^\dagger+\hat{b} \right)- \frac{\chi^2}{\omega_{20}} \hat{a}^\dagger \hat{a} \hat{a}^\dagger \hat{a}.
\end{align}
Thus, we show that the eliminated mode leads to a Kerr-like nonlinearity for the dissipative boson, which for cavity-QED systems correspond to the cavity photons.
}

\begin{figure}[!ht]
\centering
\includegraphics[width=1\columnwidth]{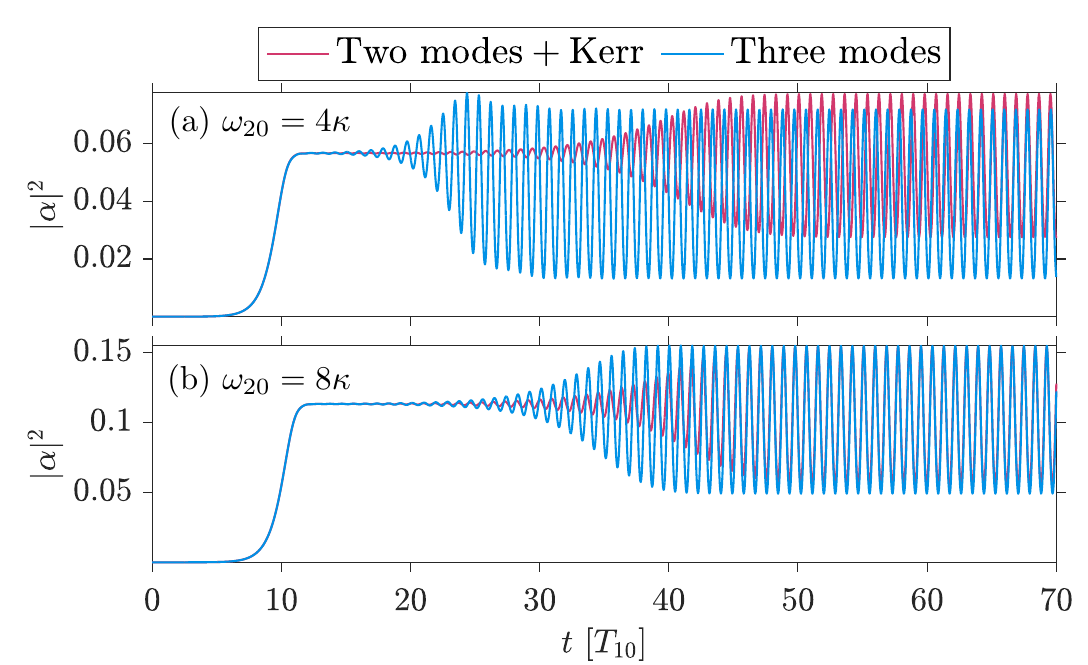}
\caption{Comparison of the mean-field dynamics according to the two-mode model with Kerr-like nonlinearity and the three-mode model for (a) $\omega_{20}=4\kappa$ and (b) $\omega_{20}=8\kappa$. The coupling strength is fixed to $\lambda = 1.09 \lambda_\mathrm{SR}$. The remaining parameters are $\kappa=\omega_p=\chi/2=\omega_{10}/2$. } 
\label{fig:2} 
\end{figure}

{In Fig.~\ref{fig:2}, we compare the mean-field dynamics for the three-mode model and the two-mode model with the Kerr-like nonlinearity obtained by numerically solving Eqs.~\eqref{eq:eom3mode} and \eqref{eq:eomhp}, respectively. Here, we fixed the coupling strength to $\lambda = \lambda_\mathrm{SR}$ with $\lambda_\mathrm{SR}$ as the critical point signalling the instability of the trivial fixed point $\alpha=\beta=\gamma=0$.
For larger $\omega_{20}$ exemplified in Fig.~\ref{fig:2}(b), the quantitative agreement between the two models improves since the adiabatic elimination of the $c$-mode hinges on the assumption that $\omega_{20} \ll \omega_{10},\omega_p$. Nevertheless, we find qualitative agreement for the type of response (i.e., an LC phase) for $\omega_{20} = 4 \kappa$, a parameter choice motivated by the experiment that will be discussed later. 
}

\subsection{Bifurcation theory}
{
To understand the nature of different critical transitions in the system as shown in Fig.~\ref{fig:1}(d), we employ a stability matrix analysis for fixed points of the semiclassical EOM Eq.~\eqref{eq:eomhp}. 
The EOM for the three-mode model prior to the adiabatic elimination, Eq.~\eqref{eq:eom3mode}, can be recasted into $\partial_t \mathbf{X}=\mathbf{F}(\mathbf{X})$ with $\mathbf{X}=\left\{\alpha,\alpha	*,\beta,\beta^*,\gamma,\gamma^*\right\}$. Numerically solving for the equilibrium or fixed points  $\mathbf{X}_0=\left\{\alpha_0,\alpha^*_0,\beta_0,\beta^*_0,\gamma_0,\gamma^*_0\right\}$, such that $\mathbf{F}(\mathbf{X}_0)=0$, and linearising the EOM around those, we obtain a linearised set of EOM given by
\begin{equation}
\label{eq:J}
\partial_t \delta_{\mathbf{X}}= \mathbf{J}_0 \delta_{\mathbf{X}},
\end{equation}
where $\delta_{\mathbf{X}}=\left( \mathbf{X}-\mathbf{X}_0 \right) $ and $\mathbf{J}_0 =\left. \frac{\partial \mathbf{F}(\mathbf{X})}{\partial \mathbf{X}}\right|_{\mathbf{X}_0}$ is the Jacobian stability matrix. In the case of the three-mode model, the Jacobian matrix is
\begin{widetext}
\begin{align}
\mathbf{J}_0 =\left. \frac{\partial \mathbf{F}(\mathbf{X})}{\partial \mathbf{X}}\right|_{\mathbf{X}_0}= \begin{pmatrix}
-i \left[\omega_p+\chi \left( \gamma_0+\gamma^*_0 \right) \right]-\kappa & 0 & -i \lambda & -i\lambda & -i\chi \alpha_0 & -i\chi \alpha_0 \\
0 & i \left[\omega_p+\chi \left( \gamma_0+\gamma^*_0 \right) \right]-\kappa & i \lambda & i \lambda  & i\chi \alpha^*_0 & i\chi \alpha^*_0\\
-i \lambda & -i \lambda & -i \omega_{10} & 0 & 0 & 0\\
i \lambda & i \lambda & 0 & i \omega_{10} & 0 & 0\\
-i \chi \alpha^*_0 & -i \chi \alpha_0 & 0 & 0 & -i\omega_{20} & 0\\
i \chi \alpha^*_0 & i \chi \alpha_0 & 0 & 0 & 0 & i \omega_{20}
\end{pmatrix}.
\end{align}
\end{widetext}
}

{
For the two-mode model with Kerr-like nonlinearity described by Eq.~\eqref{eq:eomhp}, applying a similar linearisation leads to the following Jacobian stability matrix
}
\begin{align}
\label{mat:ja}
\mathbf{J}_0 = \begin{pmatrix}
\omega_p-i \kappa -\frac{4 |\alpha_0|^2\chi^2}{\omega_{20}} & -\frac{2 \alpha^2_0 \chi^2}{\omega_{20}} & \lambda & \lambda\\
\frac{2 (\alpha^*_0)^2 \chi^2}{\omega_{20}} & \frac{4 |\alpha_0|^2 \chi^2}{\omega_{20}}-i \kappa -\omega_p & -\lambda & -\lambda \\
\lambda & \lambda & \omega_{10} & 0\\
-\lambda & -\lambda & 0 & \omega_{10}
\end{pmatrix},
\end{align}
where $\mathbf{X}_0=  \left\{ \alpha_0,\alpha^*_0,\beta_0,\beta^*_0  \right\} $. 

\begin{figure}[!hb]
\centering
\includegraphics[width=1.0\columnwidth]{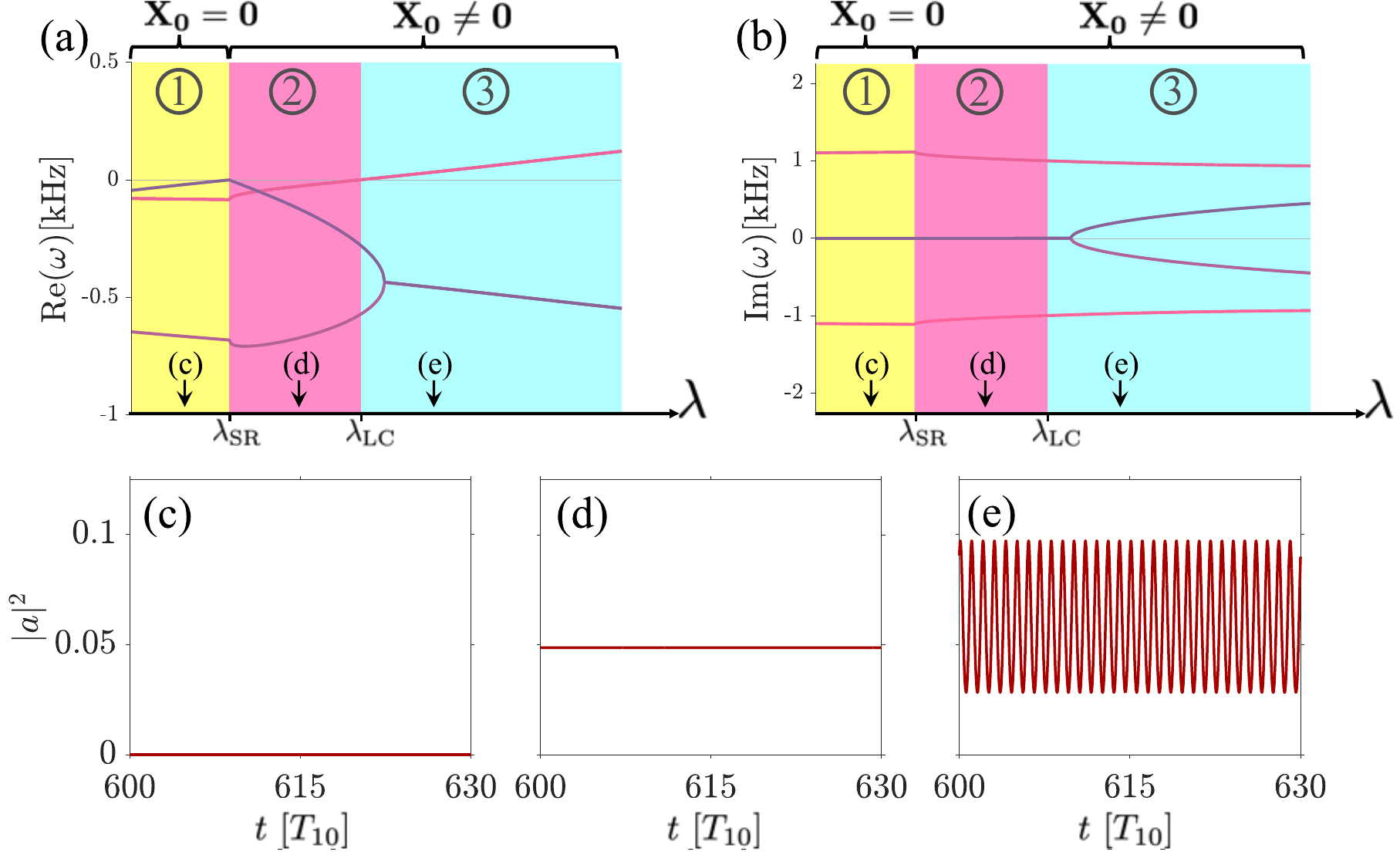}
\caption{Spectrum and dynamics for the two-mode model with a Kerr-like nonlinearity. (a),(b) Real and imaginary part of the eigenvalues obtained from numerically diagonalizing the stability matrix.  Shaded background indicates the different phases: normal phase (yellow (1)), superradiant phase (pink (2)) and limit cycle phase (light blue (3)). (c)-(e) corresponding light field dynamics in each shaded background.}
\label{fig:3} 
\end{figure}

The solutions of Eq.~(\ref{eq:J}) can be written as a superposition of $\mathrm{exp}\left(\omega_i t\right)$ with $\omega_i$ being the eigenvalues (EVs) of the Jacobian matrix (\ref{mat:ja}). 
The fixed points are only stable if the real part of all the EVs are negative $\mathrm{Re}\left( \omega_i \right)<0~\forall~i$ \cite{kosior2022}. 
{Focusing on the two-mode model with Kerr-like nonlinearity described by the stability matrix in Eq.~\eqref{mat:ja}}, we present the real and imaginary parts of the EVs of $\mathbf{J}_0$ using the appropriate fixed points in Figs.~\ref{fig:3}(a) and \ref{fig:3}(b), respectively. 
Exemplary dynamics of the occupation $|\alpha|^2$, which in the atom-cavity platform correponds to the photon number in the cavity, for various phases are shown in Figs.~\ref{fig:3}(c)-\ref{fig:3}(e), which we obtain by solving the mean-field EOM of Eq.~\eqref{eq:eom}. The parameters are $\kappa=\omega_p=\chi/2=\omega_{10}/2=\omega_{20}/4$. We choose $\omega_p=\kappa$ since this corresponds to the weakest light-matter coupling needed to enter the SR phase $\lambda_\mathrm{SR}$ for fixed $\kappa$, which can be inferred by setting $\partial \lambda_\mathrm{SR} / \partial \omega_p =0$ and solving for $\omega_p$. 
In Figs.~\ref{fig:3}(c)-\ref{fig:3}(e), we use the NP as the initial state and linearly increase the light-matter coupling strength to its finale value within $\approx 150~T_\mathrm{10}$,  where $T_{10} = 2\pi/\omega_{10}$. To rule out transient behaviour we only present the dynamics after $600~T_\mathrm{10}$.

In Fig.~\ref{fig:3}(a), for $\lambda < \lambda_\mathrm{SR}$, we use the fixed point $\alpha=\beta=\gamma=0$ corresponding to the so-called normal phase (NP) and find that, as expected, all $\mathrm{Re}\left( \omega_i \right)$ are negative, thereby confirming its stability. The dynamics of the NP is depicted in Fig.~\ref{fig:3}(c) confirming a steady-state value of $|\alpha|^2 = 0$. Above the critical point $\lambda_\mathrm{SR}$, the NP fixed point acquires an EV with a positive real part {(see Appendix \ref{sec:spec})}, which suggests an instability of this fixed point manifesting itself as a phase transition from the NP to a superradiant (SR) phase in the context of the Dicke model. This transition is a supercritical pitchfork bifurcation, meaning that the real and imaginary parts of the two relevant EVs are zero at $\lambda_\mathrm{SR}$. In the SR region highlighted by the pink area in Figs.~\ref{fig:3}(a) and \ref{fig:3}(b), we obtain two new fixed points corresponding to the pair of symmetry broken states in the SR phase. Expanding around the SR fixed points, indeed, we find that they are stable in the SR region as their $\mathrm{Re}\left( \omega_i \right)$ are all negative. The time evolution in the SR phase depicted in Fig.~\ref{fig:3}(d) shows a constant photon occupation.

Our model exhibits a second critical point $\lambda_\mathrm{LC}$ at which a supercritical Hopf bifurcation occurs, which signals an instability towards a formation of a LC. 
The LC region in Figs.~\ref{fig:3}(a) and \ref{fig:3}(b) is depicted in light blue. In contrast to the pitchfork bifurcation, the relevant EVs cross the real axis, while their imaginary parts are nonzero. An exemplary LC dynamics is shown in Fig.~\ref{fig:3}(e). The photon number oscillates at a frequency given by the imaginary part of the corresponding EVs. For a Hopf bifurcation, the oscillation amplitude of the LCs increases as $\sqrt{\lambda-\lambda_\mathrm{LC}}$ \cite{Strogatz2000} {and we show that the LCs in this paper follows this scaling behaviour in Appendix \ref{sec:lcdyn}.} In Fig.~\ref{fig:1}(e), we present the photon dynamics in the phase space spanned by the real and imaginary parts of the photon field. It demonstrates how the system approaches the stable LC orbit starting from an SR phase marked by the gray diamond. The red curves represent transient oscillations and the blue lines correspond to the final LC orbit. {In Appendix \ref{sec:lcdyn}, we show the trajectories of different initial states converging to the same LC orbit, which is a characteristic feature of a limit cycle attractor}.

{Before we discuss the experimental results, we briefly mention further features of the LCs found in the smaller islands in the phase diagram as shown in Appendix \ref{sec:lcdyn}}. Here, an LC phase oscillates between the two fixed points while accumulating a phase in the photon field at integer steps of $\pi$. This suggests the presence of a particle current similar to the self-oscillating pumping reported in Ref.~\cite{Dreon2022,Nie2023}. We note that this type of LC is not due to a Hopf bifurcation as the scaling of the LC oscillations amplitude is approximately constant with $\lambda$, and therefore inconsistent with the Hopf bifurcation scaling $\sqrt{\lambda-\lambda_\mathrm{LC} }$. {Exemplary dynamics of the photon number and the phase winding are presented in Appendix \ref{sec:lcdyn}.}

\begin{figure*}[!ht]
\centering
\includegraphics[width=2\columnwidth]{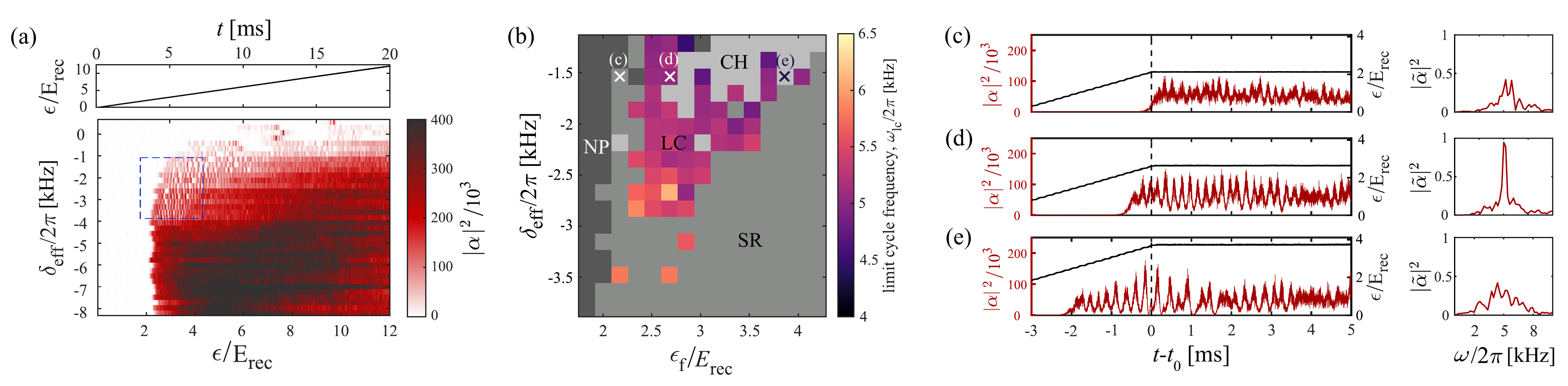}
\caption{Experimental data for a red-detuned CTC or LC. (a) Pump strength protocol (top) and the intracavity photon number $|\alpha|^2$  (bottom) for varying the effective cavity field frequency $\delta_\mathrm{eff}$ and linearly ramped pump strength $\epsilon$ in units of the recoil energy $E_\mathrm{rec}$. The blue dashed box depicts the area of the parameter space further analysed in (b). (b) Phase diagram for varying $\delta_\mathrm{eff}$ and final pump strength $\epsilon_f$. For each plaquette, we linearly ramp the pump strength to $\epsilon_f$ while keeping $\delta_\mathrm{eff}$ constant. We categorize the normal phase (NP), superradiant (SR) phase, chaotic phase (CH), and the limit cycle (LC) phase. The colour represents the dominant oscillation frequency in the LC region. (c)-(e) Exemplary dynamics and the resulting power spectrum $\tilde{\alpha}(\omega)$ of the SR, LC, and chaotic phase marked by crosses in (b). $t_0$ is the time at which the pump strength is fully ramped up and is kept constant.}
\label{fig:4} 
\end{figure*} 

\section{Atom-cavity implementation}\label{sec:expt}

\subsection{Mapping from the atom-cavity system}

{In the following, we focus on a specific implementation using an atom-cavity setup \cite{Baumann2010, Klinder2015} as sketched in Fig.~\ref{fig:1}(c). In the context of LCs and CTCs found in the atom-cavity platform \cite{Kongkhambut2022}, an important implication of our theory is the possibility of observing LCs even for attractive pump field potentials as demonstrated in Fig.~\ref{fig:3}, which is consistent with the predictions in Ref.~\cite{Gao2023}}. We now show that this is indeed the case for an atom-cavity system operating in the recoil-resolved or good cavity limit $\kappa \sim \omega_{01}$ \cite{Kessler2014,Klinder2016}.  {The details of the derivation for mapping the atom-cavity Hamiltonian onto the effective model Eq.~\eqref{eq:eom} can be found in Appendix \ref{sec:atomcavity}}. In what follows, we will simply sketch the crucial steps. We start from a two-dimensional many-body Hamiltonian \cite{Ritsch2013,Cosme2019} neglecting both the trapping potential and contact interactions between the atoms. A study on the influence of inhomogeneous trapping and short-range interactions on a dissipative time crystal in an atom-cavity system reveals the persistence of the time crystalline phase \cite{Tuquero2022}. Next, we expand the atomic field operator in the basis of three momentum excitations of the BEC. The first state in the three-level model $|0\rangle$ is represented by the zero-momentum mode $|p_x,p_y\rangle =| 0 , 0 \rangle$ with an energy $E_0=0$. The second level $|1\rangle$ is given by the coherent superposition of $|\pm \hbar k ,\pm \hbar k \rangle$ momentum modes with an energy  $E_1=2 \hbar \omega_\mathrm{rec}$. The third level $|2\rangle$ corresponds to the coherent superposition of the $| 0, \pm 2 \hbar k \rangle $ momentum modes along the cavity axis with an energy $E_2=4 \hbar \omega_\mathrm{rec}$. This expansion then leads to the effective three-level model in Fig.~\ref{fig:1}(b). Similar models have been recently studied in \cite{Wolf2018,Skulte2021, Kongkhambut2021, Lin2022, Fan2023, Skulte2023, Valencia2023}. After using an $\mathrm{SU}(3)$ representation via the Schwinger boson mapping, we apply the HP approximation \cite{Wagner1975} to finally obtain Eq.~\eqref{eq:eom3mode}. 

{Now that we have shown that Eq.~\eqref{eq:eom3mode} can be obtained from the atom-cavity model, one can simply follow the adiabatic elimination discussed in Sec.~\ref{sec:theory} to get Eq.~\eqref{eq:eomhp}.
Alternatively, we can first derive a nonlinear Dicke model, which can then be approximated as the two-mode model with Kerr-like nonlinearity in the thermodynamic limit, by first adiabatically eliminating the third-level in the few-mode atom-cavity description in Appendix \ref{sec:atomcavity} prior to employing the Schwinger-boson mapping. In doing so, we only need $\mathrm{SU}(2)$ spin operators as in the standard Dicke model leading a nonlinear Dicke Hamiltonian} 
\begin{align}
\label{eq:ham}
\hat{H} =\hat{H}_\mathrm{Dicke}+\hat{H}_\mathrm{Kerr},
\end{align}
where the Dicke Hamiltonian is
\begin{align}
\label{eq:ham1}
\frac{\hat{H}_\mathrm{Dicke}}{\hbar} &= \omega_p \hat{a}^\dagger \hat{a}+\omega_{10} \sum_{\ell=1}^{N}  \sigma_\ell^z + \frac{2 \lambda}{\sqrt{N}} \sum_{\ell=1}^{N}  (\hat{a} + \hat{a}^\dagger) \sigma_\ell^x, 
\end{align}
with $\sigma_\ell^\mu$ as the individual $\mathrm{SU}(2)$ spin operators. The bosonic operators $\hat{a}$ and $\hat{a}^\dagger$ annihilate and create a photon in the quantised light mode, respectively. The Kerr-like Hamiltonian is
\begin{align}
\label{eq:ham2}
\hat{H}_\mathrm{Kerr}/ \hbar &=- \frac{\chi^2}{\omega_{20}} \hat{a}^\dagger \hat{a} \hat{a}^\dagger \hat{a} .
\end{align}
We emphasize that while we consider a third level to be the origin of the Kerr nonlinearity, {other physical processes generating this nonlinearity will result in the same phenomena, see for example Refs.~\cite{aspelmeyer2014, larson2021, Rameshti2022}}.
Introducing $j_\mu = \frac{1}{\sqrt{N}} \langle \sum_\ell \sigma_\ell^\mu \rangle$, with $\mu \in {x,y,z}$, the EOM for the spin-boson model are given by
\begin{align}
\frac{\mathrm{d} \alpha}{\mathrm{d} t} &= -i \left[ \omega_p-2 \frac{\chi^2}{\omega_{20} } |\alpha|^2-i\kappa  \right] \alpha-2 i \lambda j_x \notag \\
 \frac{\mathrm{d} j_x}{\mathrm{d} t} &= -\omega_{10} j_y \notag \\ \frac{\mathrm{d} j_y}{\mathrm{d} t} &=  \omega_{10} j_x-2 \lambda \left(\alpha+\alpha^* \right)  j_z \notag 
 \\ \frac{\mathrm{d} j_z}{\mathrm{d} t} &=  2 \lambda \left(\alpha+\alpha^* \right)  j_y
\label{eq:eom}
\end{align}
with $\omega_p$ is the photon frequency, and $\omega_{nm}$ is the level splitting between the atomic states $|n\rangle$ and $|m\rangle$, see Fig.\ref{fig:1}(b). The light-matter interaction proportional to $\lambda$ couples the atomic modes with the two lowest energies $|0\rangle$ and $|1\rangle$ as in the standard dipole approximation.  Here, we also consider a two-photon coupling between the first and third atomic levels $|0\rangle$ and $|2\rangle$, which we adiabatically eliminate to obtain the Kerr-like nonlinearity for the photonic field. The strength of this nonlinearity is controlled by $\chi$. We included the decay strength $\kappa$ in the photonic mode equation of motion, which captures the rate at which photons are emitted from the cavity.
We apply a Holstein-Primakoff (HP) transformation and include only terms up to linear order in the bosonic operator \cite{Emary2003, Skulte2021} and obtain precisely the EOM in Eq.~\eqref{eq:eomhp} 

\subsection{Experimental results}

We experimentally demonstrate the emergence of a LC phase for an attractive light-shift pump potential in the atom-cavity platform schematically shown in Fig.~\ref{fig:1}(b). We emphasise that this in contrast to the repulsive light-shift pump potential used in the theoretical prediction \cite{Piazza2015,Hans2019,Kessler2020} and experimental realisation \cite{Kongkhambut2022} of LCs in the atom-cavity system. That is, we provide here the first experimental observation of a LC or CTC for an attractive light-shift pump potential, which underpins the mechanism put forth by our generic model Eq.~\eqref{eq:eom}. That is, the transition from an SR to an LC phase can be understood as a Hopf bifurcation induced by the Kerr nonlinearity $\chi$ that gives access to a third level $|2\rangle$ beyond the usual two-level approximation that maps the atom-cavity system onto an open Dicke model. 

In our experiment, we place a BEC consisting of $N\approx 4\times 10^4$ $^{87}\mathrm{Rb}$ atoms inside a high-finesse cavity that is pumped transversely by a retro reflected laser beam, which produces a standing wave potential for the atoms. The pump wavelength used is $803.63~\mathrm{nm}$, which is red-detuned to the relevant atomic transition  at $794.98~\mathrm{nm}$. The resulting two-photon coupling strength is $\chi/2\pi \approx -6~\mathrm{kHz}$. The recoil energy $E_\mathrm{rec}/\hbar = 2\pi\times3.55~\mathrm{kHz}$ is comparable to the cavity decay rate of $\kappa = 2\pi\times 3.6~\mathrm{kHz}$. Thus, the dynamics of the light field and the atoms evolve on the same time scale and influence each other on equal footing. To identify the approximate regime of LCs, we record the photon number $|\alpha|^2$ over $5~\mathrm{ms}$ after the pump strength is slowly ramped up from zero to its final value at a rate of $0.6~\frac{E_\mathrm{rec}}{\mathrm{ms}}$. We show the corresponding results for different effective pump-cavity detuning $\delta_\mathrm{eff} \sim -\omega_p$ in Fig.~\ref{fig:4}(a). In addition to the standard NP-SR phase transition \cite{Baumann2010,Klinder2015}, we observe oscillatory behaviour for certain values of small $\rvert\delta_\mathrm{eff}\lvert$ after entering the SR phase, which is indicative of a LC phase. 

We focus on the region enclosed by the dashed blue box in Fig.~\ref{fig:4}(a). 
For these combinations of $\epsilon_f$ and $\delta_\mathrm{eff}$, we now ramp up the pump strength with the same rate as before to its desired final value, which is then kept constant, 
{see Appendix \ref{sec:exptpd} for details on the construction of the phase diagram}. 
The resulting phase diagram is shown in Fig.~\ref{fig:4}(b). Comparing the overall shape of the experimental LC regime in Fig.~\ref{fig:4}(b) and the theoretical results presented in Appendix \ref{sec:lcdyn}, we find qualitative agreement. However, we point out that the approximations applied in our theory lead to a larger area with stable LCs than in the experiment. We further note that, in the experiment, the lifetime of the LCs is limited by atom loss induced by three-body collisions, which essentially reduces the light-matter coupling $\lambda$ and non-linearity $\chi$, and by the inherent short-range interactions, which has been proposed to make the LC metastable \cite{johansen2023role}. The atom loss effectively drags the system to the bottom-left region of the phase diagram Fig.~\ref{fig:4}(b), which brings it back to the NP. In Figs.~\ref{fig:4}(c)-\ref{fig:4}(e), we present three exemplary traces of the time evolution of the photon number and their corresponding power spectra, characterizing the chaotic (CH), LC and SR phases marked in Fig.~\ref{fig:4}(b). 
Approaching the LC phase boundary from the SR phase, in the intra-cavity photon number versus time, shown in Fig.~\ref{fig:4}(c), we observe a constant population level of the cavity mode together with noise, with a power spectrum showing a broad peak around the LC-frequency, which we interpret as a precursor of the LC phase {(see Appendix \ref{sec:exptpd} for details)}. Increasing the pump strength, and thus the light-matter coupling, we observe LC dynamics with a single dominant frequency peak in the power spectrum. Increasing the pump strength even further leads to aperiodic dynamics as seen directly in the time evolution of the photon number, which exhibits a largely broadened power spectrum. 

\section{Conclusions}\label{sec:conc}
In conclusion, we have proposed a generic three-mode model and a nonlinear two-mode model that features different types of LCs in a wide range of parameters. We motivate this model as the mean-field approximation of an extended Dicke model, but emphasize that it emerges generically in a broad class of systems. The predominant type of LC in the model arises from a supercritical Hopf bifurcation. The Hamiltonian of this model can be implemented by systems that can be approximated as coupled bosonic modes using a HP transformation. We show that the existence of LCs are independent of the sign of the Kerr nonlinearity introduced by a two-photon process, which is relevant for the formation of LCs in atom-cavity systems. More specifically, it is the presence of a third mode and its coupling with the density of a dissipative bosonic mode that leads to an effective Kerr-like coupling, which introduces the nonlinearity needed for the emergence of LCs. Using an atom-cavity platform, we experimentally observe for the first time the emergence of stable LCs for attractive or red-detuned pump fields. We map out the phase diagram and find good qualitative agreement with the theoretical results. Our work puts forth a new mechanism for creating LCs and studying nonlinear dynamics in highly controllable quantum systems. We emphasise that our bosonic model Eq.~\eqref{eq:eom} and Eq.~\eqref{eq:eomhp} is not limited to atom-cavity systems and can be used in a wider class of systems involving boson-mediated interactions and three-level systems, such as cavity-magnon models \cite{Zare2022}, circuit QED \cite{Chang2020,Minganti2023}, or Rydberg platforms  \cite{wu2023,wadenpfuhl2023}.

\begin{acknowledgments}
This work was funded by the UP System Balik PhD Program (OVPAA-BPhD-2021-04), the QuantERA II Programme that has received funding from the European Union's Horizon 2020 research and innovation programme under Grant Agreement No 101017733, the Deutsche Forschungsgemeinschaft (DFG, German Research Foundation) ``SFB-925" project 170620586, and the Cluster of Excellence ``Advanced Imaging of Matter" (EXC 2056), Project No. 390715994. J.S. acknowledges support from the German Academic Scholarship Foundation. H.K. acknowledges funding by the state of North Rhine-Westphalia through the EIN Quantum NRW program. JGC thanks Ryo Hanai for valuable insights and discussions. 
\end{acknowledgments}

\appendix

\section{Spectrum using $\mathbf{X}_0=0$}\label{sec:spec}
\begin{figure}[!ht]
\centering
\includegraphics[width=1\columnwidth]{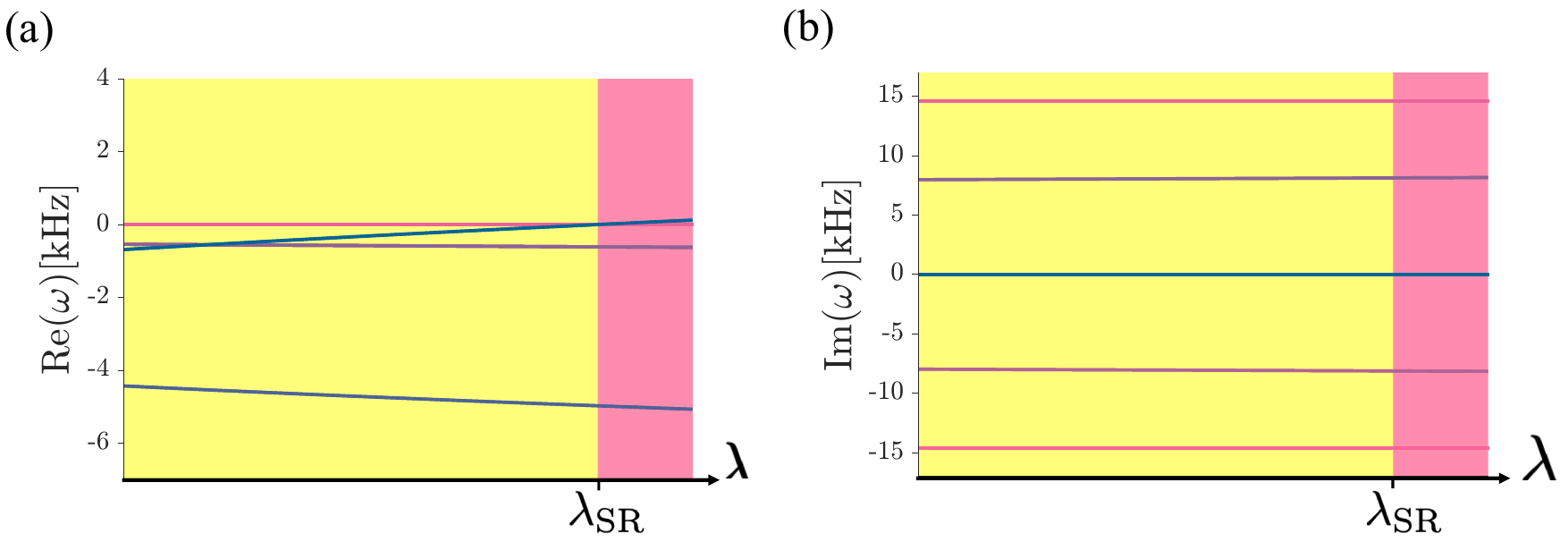}
\caption{(a) Real and (b) imaginary parts of the eigenvalues of the stability matrix for the boson-mediated three-level system with the fixed point $\mathbf{X}_0=0$. The fixed point $\mathbf{X}_0=0$ becomes unstable for $\lambda > \lambda_\mathrm{SR}$.}
\label{sfig:1} 
\end{figure}

We present in Fig.~\ref{sfig:1} the spectrum of the Jacobian using the NP fixed point $\mathbf{X}_0=0$ for different $\lambda$. For $\lambda > \lambda_\mathrm{SR}$, the system acquires an EV with $\mathrm{Re}(\omega_i) > 0$ as seen in one of the blue curves in Fig.~\ref{sfig:1}(a). Hence, the NP fixed point becomes unstable.

\section{Limit cycle dynamics}\label{sec:lcdyn}

\subsection{Amplitude scaling of the LC transition}
\begin{figure}[!ht]
\centering
\includegraphics[width=0.6\columnwidth]{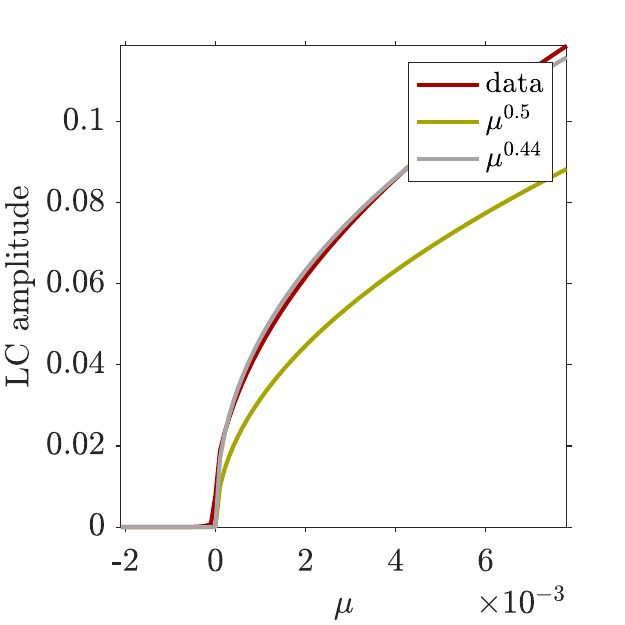}
\caption{Scaling of the LC oscillations amplitude on $\mu = \lambda-\lambda_\mathrm{LC}$.}
\label{sfig:2} 
\end{figure}

We show the characteristic scaling of the LC oscillations amplitude as we cross the SR-LC phase transition. For LC stemming from Hopf bifurcations, the amplitude of the LC scales with $\mu=\lambda-\lambda_\mathrm{LC}$ as $\mu^{1/2}$ for $\lambda> \lambda_\mathrm{LC}$ \cite{Strogatz2000}. We present the scaling as a function of $\mu$ in Fig.~\ref{sfig:2}. The scaling agrees well with the theoretical prediction. We attribute deviations from the expeted $\mu^{1/2}$ behaviour to asymmetry in the LC dynamics per cycle. We further note that, while we are very close to the phase transition, numerically its not feasible to zoom in further for even smaller $\mu$ as the systems takes too long to approach the corresponding stable LC phase. This could also further lead to deviations from the expected scaling.

\subsection{Different initial states}
\begin{figure}[!ht]
\centering
\includegraphics[width=0.8\columnwidth]{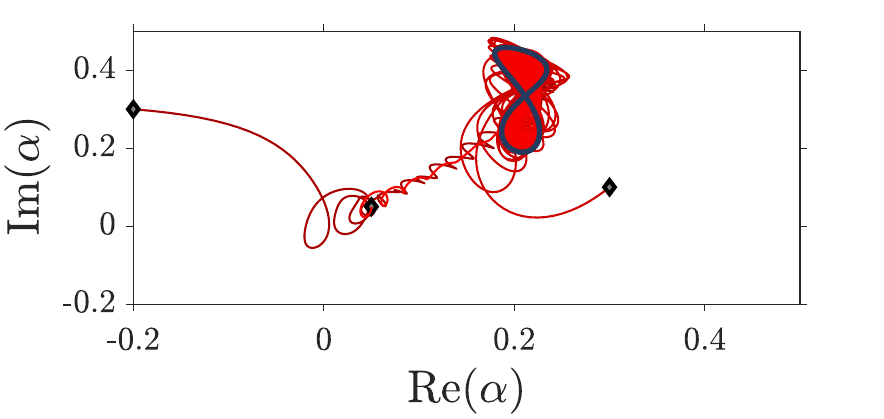}
\caption{Dynamics in the phase space of the photon mode $\alpha$ for different initial states (gray diamonds) and their approach towards the limit cycle (blue line).}
\label{sfig:4} 
\end{figure}

We present the dynamics of the transient behaviour of the light-field during the transition from a random initial state towards the stable LC orbit. We find that independent of the initial state, the steady state dynamics is the same LC orbit, which is the defining behaviour of LC dynamics. We demonstrate this for three different initial states in Fig.~\ref{sfig:4}.

\subsection{Phase diagram for blue- and red-detuned pump frequencies}

\begin{figure}[!htpb]
\centering
\includegraphics[width=\columnwidth]{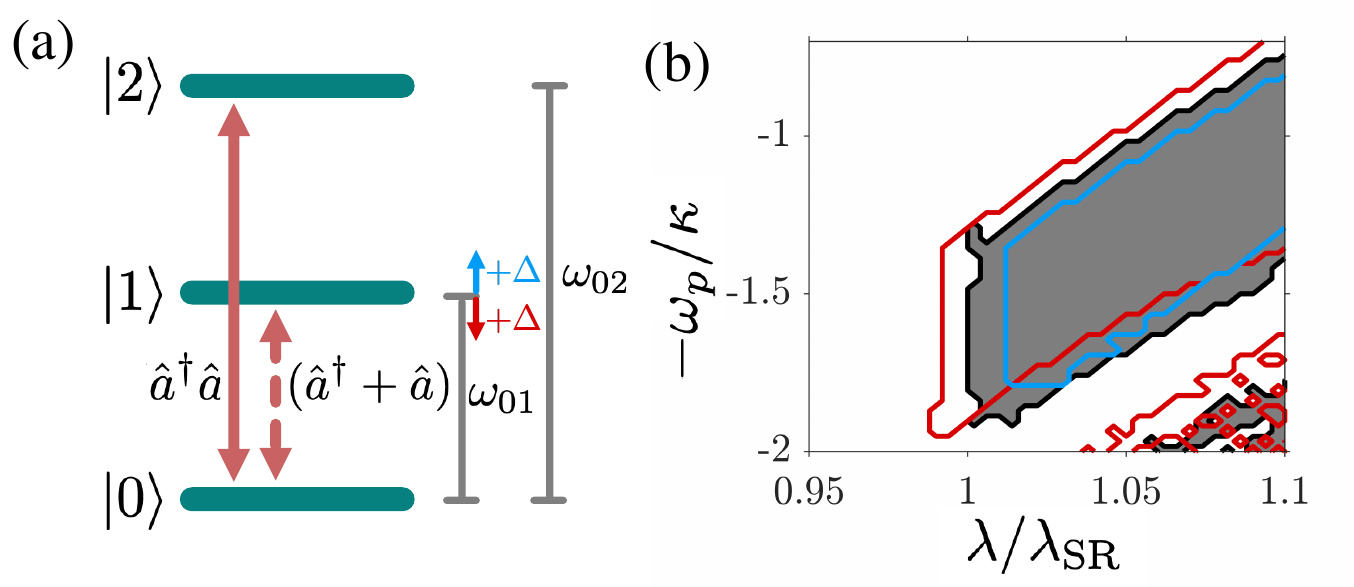}
\caption{(a) Three-level model coupled to a single light mode including single-photon coupling with strength $\lambda$ and two-photon coupling with strength $\chi$. (b) Phase diagram of the three-level model in (a) and Fig1.~(c) for varying interaction strengths $\lambda$ and photon
frequencies $\omega_\mathrm{p}$. Areas enclosed by lines denote a stable limit cycle regime. Black denotes results for the system in (a) without the level dressing of the transverse pump field, while blue and red correspond to blue- and red-detuned pump frequencies relative to $\omega_{01}$, respectively.}
\label{sfig:5} 
\end{figure}

Solving the corresponding equations of motion (EOM) allows us to construct the phase diagram in Fig.~\ref{sfig:5}(b) for different combinations of $\omega_p$ and $\lambda$. The vertical axis is displayed as $-\omega_p/\kappa$ since this is related to the effective detuning between the pump and cavity fields, which is chosen to be negative in atom-cavity experiments. Unless indicated otherwise, we fix the two-photon coupling strength to $\chi / \kappa \approx 4$, which is a typical value in atom-cavity experiments as we will show later. In Fig.~\ref{fig:1}(d), we only highlight the regimes with stable LCs, although we note that the system also hosts a transition between a normal (NP) and a superradiant (SR) phase at a critical light-matter interaction $\lambda_\mathrm{SR}=\sqrt{(\kappa^2+\omega^2)\,\omega_{10}/4\omega}$. We also find chaotic phases marked by irregular dynamics of the photon occupation. We classify a periodic dynamics as a limit cycle if the steady-state amplitude of the oscillations satisfy $\mathrm{max}(|\alpha|^2)/\mathrm{mean}(|\alpha|^2)>0.02$ and the long-time standard deviation of the oscillation peaks is $\sigma_{|\alpha|^2}<0.025$.  These LC phases are equivalent to the CTCs observed in Ref.~\cite{Kongkhambut2022}.

The atomic levels can be dressed by the pump field leading to energy shifts $\Delta=\mathrm{sign}(\chi)\epsilon \omega_\mathrm{rec}/4 =\mathrm{sign}(\chi) \lambda^2/ 8 \omega_{\mathrm{rec}}  \chi $, where $\epsilon$ is the intensity of the pump field and $\omega_\mathrm{rec}$ is the associated recoil frequency. Neglecting the pump laser dressing in the atom-cavity system means that the frequency splitting between the $|0\rangle$ and $|1\rangle$ is simply given by the recoil frequency, i.e., $\omega_{01} = 2\omega_\mathrm{rec}$. A more accurate theory includes such dressing \cite{Skulte2021,Kongkhambut2021}, which then adjusts the frequency of $|1\rangle$ depending on the sign of the frequency shift per single atom $U_0$ as depicted in Fig.~\ref{sfig:5}(a).

In Fig.~\ref{sfig:5}(b) we find two distinct regimes, wherein LCs can emerge. For small $\omega_p$ and $\lambda$, we find a large area in the phase diagram hosting LCs. In contrast for large $\omega_p$ and $\lambda$, we find smaller disconnected islands of LCs. We note the energy shift due to the pump dressing $\Delta$ simply moves the LC regions without changing their overall shape in the phase diagram. Therefore, our results suggest that the emergence of the LC phase in atom-cavity systems \cite{Piazza2015,Hans2019,Kongkhambut2022} does not hinge on the repulsive nature of the pump $\Delta > 0$. Instead, the effective Kerr-like nonlinearity $\chi$ that couples the lowest energy mode to a new third mode is the crucial ingredient for the existence of the LCs or CTCs.

\subsection{Phase-winding limit cycles}
\begin{figure}[!ht]
\centering
\includegraphics[width=\columnwidth]{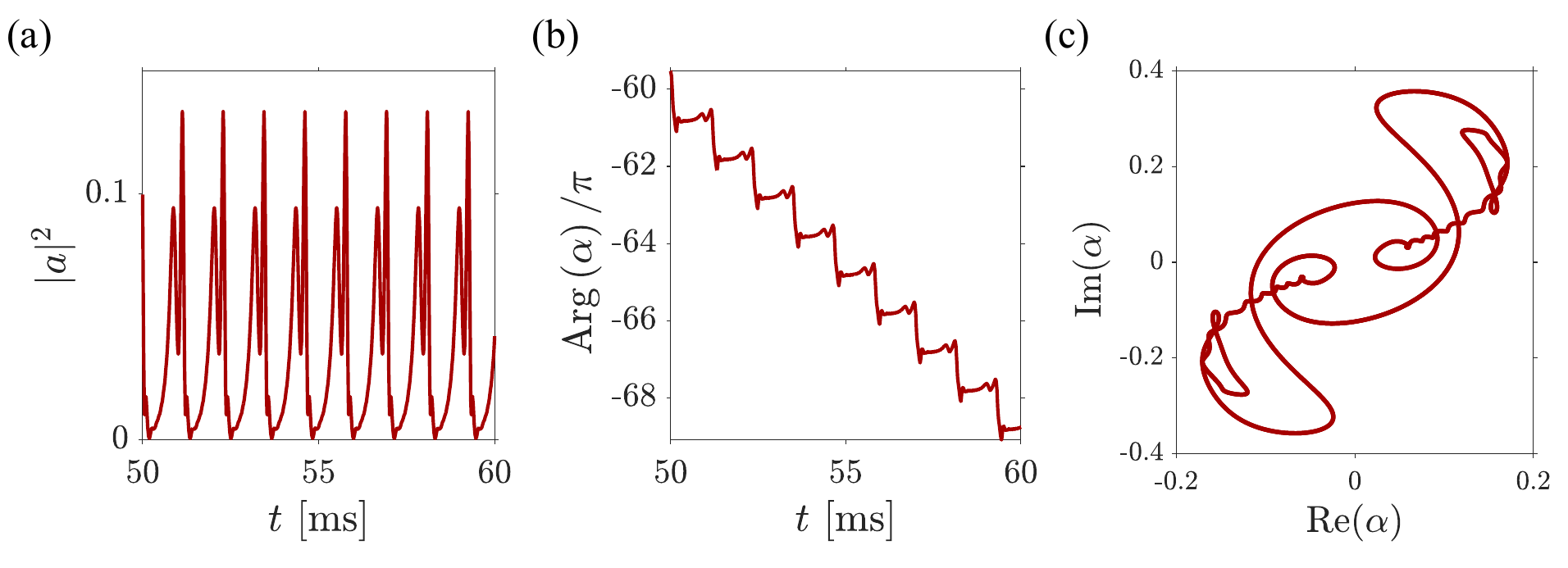}
\caption{(a) Light-field dynamics of the photon number, (b) corresponding unwrapped phase of $\alpha$, and (c) phase space dynamics.The parameters are $\omega_p/\kappa =1.9$ and $\lambda/\lambda_\mathrm{SR}=1.02$.}
\label{sfig:3} 
\end{figure}

For the type of LC found in the small islands shown in Fig.~\ref{sfig:5}, we present the photon number dynamics and the corresponding unwrapped phase in Fig.~\ref{sfig:3}. We find that, in contrast to the LCs discussed in the main text, these LCs pick up a phase of $\pi$ during each cycle.

\section{Mapping from the atom-cavity Hamiltonian}\label{sec:atomcavity}
We start from the many-body Hamiltonian describing a transversely pumped BEC inside a high-finesse cavity \cite{Ritsch2013,Kongkhambut2022}
\begin{widetext}
\begin{align}
\hat{H}/\hbar &= - \delta_C \hat{a}^\dagger \hat{a} + \int dy dz \Psi^\dagger (y,z) \left[ -\frac{\hbar}{2m} \nabla^2+\mathrm{sign}\left(U_0 \right)\omega_\mathrm{rec}\epsilon_p \cos^2(ky) \right] \Psi(y,z)\\ &+ \Psi^\dagger\left( U_0  \hat{a}^\dagger \hat{a} \cos^2(kz) -\sqrt{\omega_\mathrm{rec}|U_0|\epsilon_p}\cos(ky)\cos(kz)(\hat{a}^\dagger+\hat{a})  \right) \Psi(y,z),
\end{align}
\end{widetext}
where the pump (cavity) axis is along the $y$($z$)-direction, $\epsilon_p$ is the pump strength, $U_0$ is the maximum light-shift per atom, $\omega_\mathrm{rec}$ is the recoil frequency, and $\delta_C$ is the detuning between the pump and cavity frequencies. We expand our field operator as
\begin{equation}
\Psi = \psi_0 c_0+\psi_1 c_1+\psi_2 c_2
\end{equation}
with
\begin{align}
\psi_0 &= 1 \\
\psi_1 &= 2 \cos(ky) \cos(kz) \\
\psi_2 &= \sqrt{2} \cos(2kz)~.
\end{align}
With this, we obtain the effective Hamiltonian
\begin{align}
H/\hbar &=-\delta_\mathrm{eff} \hat{a}^\dagger \hat{a} +\omega_\mathrm{rec}\epsilon_p/2+ \left(\omega_{10}+\Delta\right)c^\dagger_1 c_1+ \omega_{20}c^\dagger_2c_2 \\ \nonumber
&+ \frac{U_0}{2}\hat{a}^\dagger \hat{a} \left(  \frac{1}{2}c^\dagger_1 c_1+(c^\dagger_2c_0+c^\dagger_0c_2)/\sqrt{2} \right) \\ \nonumber
&+\frac{\lambda}{\sqrt{N}} (\hat{a}^\dagger+\hat{a}) \left[ (c^\dagger_0c_1+c^\dagger_1c_0) +(c^\dagger_1c_2+c^\dagger_2c_1)\sqrt{2} \right]~,
\end{align}
where $\lambda/\sqrt{N}=-\omega_\mathrm{rec}|U_0|\epsilon_\mathrm{p}/2$, $\delta_\mathrm{eff}= \delta_C-U_0 N/2= \delta_C-U/2$, $\Delta=\mathrm{sign}(\chi)\epsilon \omega_\mathrm{rec}/4$, $\omega_{10}=\omega_\mathrm{rec}$ and $\omega_{20}=2 \omega_\mathrm{rec}$. In the following we assume that the the lowest mode is highly occupied. This means we neglect the term  $\frac{U_0}{4}a^\dagger a c^\dagger_1 c_1$ and $\frac{\lambda}{\sqrt{N}} (a^\dagger+a) (c^\dagger_1c_2+c^\dagger_2c_1)\sqrt{2}$ as these scale as $1/N$ compared to the other terms and we are interested in the limit $N \gg 1$. We further drop constant energy shifts of the Hamiltonian. Thus, the simplified atom-cavity Hamiltonian is
\begin{align}
&H/\hbar =-\delta_\mathrm{eff} \hat{a}^\dagger \hat{a}+ \left(\omega_{10}+\Delta\right)c^\dagger_1 c_1+ \omega_{20}c^\dagger_2c_2 \\ \nonumber
&+ \frac{\sqrt{2}U_0}{4}\hat{a}^\dagger \hat{a} \left( c^\dagger_2c_0+c^\dagger_0c_2 \right) +\frac{\lambda}{\sqrt{N}} (\hat{a}^\dagger+\hat{a}) \left(c^\dagger_0c_1+c^\dagger_1c_0\right).
\end{align}
Next, we map the three atomic modes to $\mathrm{SU}(3)$ spins using the Schwinger-Boson mapping \cite{Skulte2021} and obtain
\begin{align}
H/\hbar &=-\delta_\mathrm{eff} \hat{a}^\dagger \hat{a}+ \left(\omega_{10}+\Delta\right)\hat{J}^{01}_z+ \omega_{20}\hat{J}^{02}_z \\ \nonumber 
&+ \frac{\sqrt{2}U_0}{4}\hat{a}^\dagger \hat{a} \hat{J}^{02}_x+\frac{\lambda}{\sqrt{N}} (\hat{a}^\dagger+\hat{a})\hat{J}^{01}_x~.
\end{align}
Finally, we use the Holstein-Primakoff representation given by \cite{Wagner1975}
\begin{align}
\hat{J}^{01}_z &= \hat{b}^\dagger\hat{b}-N/2, \quad \hat{J}^{01}_+ = \hat{b}^\dagger~\sqrt{N-\left( \hat{b}^\dagger  \hat{b}+\hat{c}^\dagger \hat{c} \right)} , \\ \nonumber
\hat{J}^{01}_- &=  \sqrt{N-\left( \hat{b}^\dagger  \hat{b}+\hat{c}^\dagger \hat{c} \right)}~\hat{b},\\ \nonumber
\hat{J}^{02}_z &= \hat{c}^\dagger\hat{c}-N/2, \quad \hat{J}^{02}_+ = \hat{c}^\dagger~\sqrt{N-\left(  \hat{b}^\dagger  \hat{b}+\hat{c}^\dagger \hat{c} \right)}, \\ \nonumber
\hat{J}^{02}_- &=  \sqrt{N-\left(  \hat{b}^\dagger  \hat{b}+\hat{c}^\dagger \hat{c} \right)}~  \hat{c},
\end{align}
and by retaining terms only up to the lowest order in $N$, with $\omega_p = - \delta_\mathrm{eff}$ and $\chi=\sqrt{2}/4U_0 N$, we obtain Eq.~\eqref{eq:H}.
\vspace{1cm}

\section{Construction of the experimental phase diagram}\label{sec:exptpd}
To construct the phase diagram from the experimental data, we consider the following three quantities to distinguish between the various phases. We consider the average photon number $\overline{|\alpha|^2}$, the standard derivation of the fluctuating photon number divided by the mean photon number $\sigma_{\overline{|\alpha|^2}}=\sigma_{|\alpha|^2}/ \overline{|\alpha|^2}$, and the so-called crystalline fraction $\xi$ \cite{Kessler2021,Kongkhambut2022}, which is defined via the amplitude of a Gaussian fit around the LC peak in the Fourier transform of the dynamics of the photon number. 

We classify trajectories with less then $2 \times 10^3$ photons on average detected to be in the normal phase. To further distinguish between the SR phase, limit cycle phase and the chaotic phase, we use the following criteria. If the average photon number is larger then  $\overline{|\alpha|^2}>2 \times 10^3$ and  $\xi \leq 1/e \times \mathrm{max}\left( \xi \right)$, then the system is classified to be in the SR phase. If the average photon number is larger then  $\overline{|\alpha|^2}>2 \times 10^3$ and $\sigma_{\overline{|\alpha|^2}}>0.55$, it is in the chaotic/aperiodic phase. Trajectories not falling into one of the previous cases are identified as LC phases.

\bibliography{references}

\providecommand{\noopsort}[1]{}\providecommand{\singleletter}[1]{#1}%
\begin{thebibliography}{52}%
\makeatletter
\providecommand \@ifxundefined [1]{%
 \@ifx{#1\undefined}
}%
\providecommand \@ifnum [1]{%
 \ifnum #1\expandafter \@firstoftwo
 \else \expandafter \@secondoftwo
 \fi
}%
\providecommand \@ifx [1]{%
 \ifx #1\expandafter \@firstoftwo
 \else \expandafter \@secondoftwo
 \fi
}%
\providecommand \natexlab [1]{#1}%
\providecommand \enquote  [1]{``#1''}%
\providecommand \bibnamefont  [1]{#1}%
\providecommand \bibfnamefont [1]{#1}%
\providecommand \citenamefont [1]{#1}%
\providecommand \href@noop [0]{\@secondoftwo}%
\providecommand \href [0]{\begingroup \@sanitize@url \@href}%
\providecommand \@href[1]{\@@startlink{#1}\@@href}%
\providecommand \@@href[1]{\endgroup#1\@@endlink}%
\providecommand \@sanitize@url [0]{\catcode `\\12\catcode `\$12\catcode
  `\&12\catcode `\#12\catcode `\^12\catcode `\_12\catcode `\%12\relax}%
\providecommand \@@startlink[1]{}%
\providecommand \@@endlink[0]{}%
\providecommand \url  [0]{\begingroup\@sanitize@url \@url }%
\providecommand \@url [1]{\endgroup\@href {#1}{\urlprefix }}%
\providecommand \urlprefix  [0]{URL }%
\providecommand \Eprint [0]{\href }%
\providecommand \doibase [0]{https://doi.org/}%
\providecommand \selectlanguage [0]{\@gobble}%
\providecommand \bibinfo  [0]{\@secondoftwo}%
\providecommand \bibfield  [0]{\@secondoftwo}%
\providecommand \translation [1]{[#1]}%
\providecommand \BibitemOpen [0]{}%
\providecommand \bibitemStop [0]{}%
\providecommand \bibitemNoStop [0]{.\EOS\space}%
\providecommand \EOS [0]{\spacefactor3000\relax}%
\providecommand \BibitemShut  [1]{\csname bibitem#1\endcsname}%
\let\auto@bib@innerbib\@empty
\bibitem [{\citenamefont {Walls}\ and\ \citenamefont
  {Milburn}(2007)}]{walls2007}%
  \BibitemOpen
  \bibfield  {author} {\bibinfo {author} {\bibfnamefont {D.}~\bibnamefont
  {Walls}}\ and\ \bibinfo {author} {\bibfnamefont {G.}~\bibnamefont
  {Milburn}},\ }\href {https://books.google.de/books?id=MiP9qhruN68C} {\emph
  {\bibinfo {title} {{Quantum Optics}}}},\ SpringerLink: Springer e-Books\
  (\bibinfo  {publisher} {Springer Berlin Heidelberg},\ \bibinfo {year}
  {2007})\BibitemShut {NoStop}%
\bibitem [{\citenamefont {Dicke}(1954)}]{Dicke1954}%
  \BibitemOpen
  \bibfield  {author} {\bibinfo {author} {\bibfnamefont {R.~H.}\ \bibnamefont
  {Dicke}},\ }\bibfield  {title} {\bibinfo {title} {{Coherence in Spontaneous
  Radiation Processes}},\ }\href {https://doi.org/10.1103/PhysRev.93.99}
  {\bibfield  {journal} {\bibinfo  {journal} {Phys. Rev.}\ }\textbf {\bibinfo
  {volume} {93}},\ \bibinfo {pages} {99} (\bibinfo {year} {1954})}\BibitemShut
  {NoStop}%
\bibitem [{\citenamefont {Kirton}\ \emph {et~al.}(2019)\citenamefont {Kirton},
  \citenamefont {Roses}, \citenamefont {Keeling},\ and\ \citenamefont
  {Dalla~Torre}}]{DickeModel}%
  \BibitemOpen
  \bibfield  {author} {\bibinfo {author} {\bibfnamefont {P.}~\bibnamefont
  {Kirton}}, \bibinfo {author} {\bibfnamefont {M.~M.}\ \bibnamefont {Roses}},
  \bibinfo {author} {\bibfnamefont {J.}~\bibnamefont {Keeling}},\ and\ \bibinfo
  {author} {\bibfnamefont {E.~G.}\ \bibnamefont {Dalla~Torre}},\ }\bibfield
  {title} {\bibinfo {title} {{Introduction to the Dicke model: from equilibrium
  to nonequilibrium, and vice versa}},\ }\href
  {https://doi.org/10.1002/qute.201800043} {\bibfield  {journal} {\bibinfo
  {journal} {Advanced Quantum Technologies}\ }\textbf {\bibinfo {volume} {2}},\
  \bibinfo {pages} {1970013} (\bibinfo {year} {2019})}\BibitemShut {NoStop}%
\bibitem [{\citenamefont {Chan}\ \emph {et~al.}(2015)\citenamefont {Chan},
  \citenamefont {Lee},\ and\ \citenamefont {Gopalakrishnan}}]{Chan2015}%
  \BibitemOpen
  \bibfield  {author} {\bibinfo {author} {\bibfnamefont {C.-K.}\ \bibnamefont
  {Chan}}, \bibinfo {author} {\bibfnamefont {T.}~\bibnamefont {Lee}},\ and\
  \bibinfo {author} {\bibfnamefont {S.}~\bibnamefont {Gopalakrishnan}},\
  }\bibfield  {title} {\bibinfo {title} {{Limit-cycle phase in
  driven-dissipative spin systems}},\ }\href
  {https://doi.org/10.1103/PhysRevA.91.051601} {\bibfield  {journal} {\bibinfo
  {journal} {Phys. Rev. A}\ }\textbf {\bibinfo {volume} {91}},\ \bibinfo
  {pages} {051601} (\bibinfo {year} {2015})}\BibitemShut {NoStop}%
\bibitem [{\citenamefont {Owen}\ \emph {et~al.}(2018)\citenamefont {Owen},
  \citenamefont {Jin}, \citenamefont {Rossini}, \citenamefont {Fazio},\ and\
  \citenamefont {Hartmann}}]{Owen2018}%
  \BibitemOpen
  \bibfield  {author} {\bibinfo {author} {\bibfnamefont {E.}~\bibnamefont
  {Owen}}, \bibinfo {author} {\bibfnamefont {J.}~\bibnamefont {Jin}}, \bibinfo
  {author} {\bibfnamefont {D.}~\bibnamefont {Rossini}}, \bibinfo {author}
  {\bibfnamefont {R.}~\bibnamefont {Fazio}},\ and\ \bibinfo {author}
  {\bibfnamefont {M.}~\bibnamefont {Hartmann}},\ }\bibfield  {title} {\bibinfo
  {title} {{Quantum correlations and limit cycles in the driven-dissipative
  Heisenberg lattice}},\ }\href {https://doi.org/10.1088/1367-2630/aab7d3}
  {\bibfield  {journal} {\bibinfo  {journal} {New Journal of Physics}\ }\textbf
  {\bibinfo {volume} {20}},\ \bibinfo {pages} {045004} (\bibinfo {year}
  {2018})}\BibitemShut {NoStop}%
\bibitem [{\citenamefont {Colella}\ \emph {et~al.}(2022)\citenamefont
  {Colella}, \citenamefont {Kosior}, \citenamefont {Mivehvar},\ and\
  \citenamefont {Ritsch}}]{Colella2022}%
  \BibitemOpen
  \bibfield  {author} {\bibinfo {author} {\bibfnamefont {E.}~\bibnamefont
  {Colella}}, \bibinfo {author} {\bibfnamefont {A.}~\bibnamefont {Kosior}},
  \bibinfo {author} {\bibfnamefont {F.}~\bibnamefont {Mivehvar}},\ and\
  \bibinfo {author} {\bibfnamefont {H.}~\bibnamefont {Ritsch}},\ }\bibfield
  {title} {\bibinfo {title} {Open quantum system simulation of faraday's
  induction law via dynamical instabilities},\ }\bibfield  {journal} {\bibinfo
  {journal} {Physical Review Letters}\ }\textbf {\bibinfo {volume} {128}},\
  \href {https://doi.org/10.1103/PhysRevLett.128.070603}
  {10.1103/PhysRevLett.128.070603} (\bibinfo {year} {2022})\BibitemShut
  {NoStop}%
\bibitem [{\citenamefont {Buča}\ \emph {et~al.}(2022)\citenamefont {Buča},
  \citenamefont {Booker},\ and\ \citenamefont {Jaksch}}]{Berislav2022}%
  \BibitemOpen
  \bibfield  {author} {\bibinfo {author} {\bibfnamefont {B.}~\bibnamefont
  {Buča}}, \bibinfo {author} {\bibfnamefont {C.}~\bibnamefont {Booker}},\ and\
  \bibinfo {author} {\bibfnamefont {D.}~\bibnamefont {Jaksch}},\ }\bibfield
  {title} {\bibinfo {title} {{Algebraic theory of quantum synchronization and
  limit cycles under dissipation}},\ }\href
  {https://doi.org/10.21468/SciPostPhys.12.3.097} {\bibfield  {journal}
  {\bibinfo  {journal} {SciPost Phys.}\ }\textbf {\bibinfo {volume} {12}},\
  \bibinfo {pages} {097} (\bibinfo {year} {2022})}\BibitemShut {NoStop}%
\bibitem [{\citenamefont {Wu}\ \emph {et~al.}(2023)\citenamefont {Wu},
  \citenamefont {Wang}, \citenamefont {Yang}, \citenamefont {Gao},
  \citenamefont {Liang}, \citenamefont {Tey}, \citenamefont {Li}, \citenamefont
  {Pohl},\ and\ \citenamefont {You}}]{wu2023}%
  \BibitemOpen
  \bibfield  {author} {\bibinfo {author} {\bibfnamefont {X.}~\bibnamefont
  {Wu}}, \bibinfo {author} {\bibfnamefont {Z.}~\bibnamefont {Wang}}, \bibinfo
  {author} {\bibfnamefont {F.}~\bibnamefont {Yang}}, \bibinfo {author}
  {\bibfnamefont {R.}~\bibnamefont {Gao}}, \bibinfo {author} {\bibfnamefont
  {C.}~\bibnamefont {Liang}}, \bibinfo {author} {\bibfnamefont {M.~K.}\
  \bibnamefont {Tey}}, \bibinfo {author} {\bibfnamefont {X.}~\bibnamefont
  {Li}}, \bibinfo {author} {\bibfnamefont {T.}~\bibnamefont {Pohl}},\ and\
  \bibinfo {author} {\bibfnamefont {L.}~\bibnamefont {You}},\ }\href@noop {}
  {\bibinfo {title} {{Observation of a dissipative time crystal in a strongly
  interacting Rydberg gas}}} (\bibinfo {year} {2023}),\ \Eprint
  {https://arxiv.org/abs/2305.20070} {arXiv:2305.20070 [cond-mat.quant-gas]}
  \BibitemShut {NoStop}%
\bibitem [{\citenamefont {Wadenpfuhl}\ and\ \citenamefont
  {Adams}(2023)}]{wadenpfuhl2023}%
  \BibitemOpen
  \bibfield  {author} {\bibinfo {author} {\bibfnamefont {K.}~\bibnamefont
  {Wadenpfuhl}}\ and\ \bibinfo {author} {\bibfnamefont {C.}~\bibnamefont
  {Adams}},\ }\href@noop {} {\bibinfo {title} {{Emergence of synchronisation in
  a driven-dissipative hot Rydberg vapor}}} (\bibinfo {year} {2023}),\ \Eprint
  {https://arxiv.org/abs/2306.05188} {arXiv:2306.05188 [physics.atom-ph]}
  \BibitemShut {NoStop}%
\bibitem [{\citenamefont {Weis}\ \emph {et~al.}(2023)\citenamefont {Weis},
  \citenamefont {Fruchart}, \citenamefont {Hanai}, \citenamefont {Kawagoe},
  \citenamefont {Littlewood},\ and\ \citenamefont
  {Vitelli}}]{weis2023exceptional}%
  \BibitemOpen
  \bibfield  {author} {\bibinfo {author} {\bibfnamefont {C.}~\bibnamefont
  {Weis}}, \bibinfo {author} {\bibfnamefont {M.}~\bibnamefont {Fruchart}},
  \bibinfo {author} {\bibfnamefont {R.}~\bibnamefont {Hanai}}, \bibinfo
  {author} {\bibfnamefont {K.}~\bibnamefont {Kawagoe}}, \bibinfo {author}
  {\bibfnamefont {P.~B.}\ \bibnamefont {Littlewood}},\ and\ \bibinfo {author}
  {\bibfnamefont {V.}~\bibnamefont {Vitelli}},\ }\href@noop {} {\bibinfo
  {title} {Exceptional points in nonlinear and stochastic dynamics}} (\bibinfo
  {year} {2023}),\ \Eprint {https://arxiv.org/abs/2207.11667} {arXiv:2207.11667
  [nlin.CD]} \BibitemShut {NoStop}%
\bibitem [{\citenamefont {Iemini}\ \emph {et~al.}(2018)\citenamefont {Iemini},
  \citenamefont {Russomanno}, \citenamefont {Keeling}, \citenamefont
  {Schir\`o}, \citenamefont {Dalmonte},\ and\ \citenamefont
  {Fazio}}]{Iemini2018}%
  \BibitemOpen
  \bibfield  {author} {\bibinfo {author} {\bibfnamefont {F.}~\bibnamefont
  {Iemini}}, \bibinfo {author} {\bibfnamefont {A.}~\bibnamefont {Russomanno}},
  \bibinfo {author} {\bibfnamefont {J.}~\bibnamefont {Keeling}}, \bibinfo
  {author} {\bibfnamefont {M.}~\bibnamefont {Schir\`o}}, \bibinfo {author}
  {\bibfnamefont {M.}~\bibnamefont {Dalmonte}},\ and\ \bibinfo {author}
  {\bibfnamefont {R.}~\bibnamefont {Fazio}},\ }\bibfield  {title} {\bibinfo
  {title} {{Boundary Time Crystals}},\ }\href
  {https://doi.org/10.1103/PhysRevLett.121.035301} {\bibfield  {journal}
  {\bibinfo  {journal} {Phys. Rev. Lett.}\ }\textbf {\bibinfo {volume} {121}},\
  \bibinfo {pages} {035301} (\bibinfo {year} {2018})}\BibitemShut {NoStop}%
\bibitem [{\citenamefont {Ke\ss{}ler}\ \emph {et~al.}(2019)\citenamefont
  {Ke\ss{}ler}, \citenamefont {Cosme}, \citenamefont {Hemmerling},
  \citenamefont {Mathey},\ and\ \citenamefont {Hemmerich}}]{Hans2019}%
  \BibitemOpen
  \bibfield  {author} {\bibinfo {author} {\bibfnamefont {H.}~\bibnamefont
  {Ke\ss{}ler}}, \bibinfo {author} {\bibfnamefont {J.~G.}\ \bibnamefont
  {Cosme}}, \bibinfo {author} {\bibfnamefont {M.}~\bibnamefont {Hemmerling}},
  \bibinfo {author} {\bibfnamefont {L.}~\bibnamefont {Mathey}},\ and\ \bibinfo
  {author} {\bibfnamefont {A.}~\bibnamefont {Hemmerich}},\ }\bibfield  {title}
  {\bibinfo {title} {{Emergent limit cycles and time crystal dynamics in an
  atom-cavity system}},\ }\href {https://doi.org/10.1103/PhysRevA.99.053605}
  {\bibfield  {journal} {\bibinfo  {journal} {Phys. Rev. A}\ }\textbf {\bibinfo
  {volume} {99}},\ \bibinfo {pages} {053605} (\bibinfo {year}
  {2019})}\BibitemShut {NoStop}%
\bibitem [{\citenamefont {Bu{\v c}a}\ \emph {et~al.}(2019)\citenamefont {Bu{\v
  c}a}, \citenamefont {Tindall},\ and\ \citenamefont {Jaksch}}]{Buca2019}%
  \BibitemOpen
  \bibfield  {author} {\bibinfo {author} {\bibfnamefont {B.}~\bibnamefont
  {Bu{\v c}a}}, \bibinfo {author} {\bibfnamefont {J.}~\bibnamefont {Tindall}},\
  and\ \bibinfo {author} {\bibfnamefont {D.}~\bibnamefont {Jaksch}},\
  }\bibfield  {title} {\bibinfo {title} {{Non-stationary coherent quantum
  many-body dynamics through dissipation}},\ }\href
  {https://doi.org/10.1038/s41467-019-09757-y} {\bibfield  {journal} {\bibinfo
  {journal} {Nature Communications}\ }\textbf {\bibinfo {volume} {10}},\
  \bibinfo {pages} {1730} (\bibinfo {year} {2019})}\BibitemShut {NoStop}%
\bibitem [{\citenamefont {Bakker}\ \emph {et~al.}(2022)\citenamefont {Bakker},
  \citenamefont {Bahovadinov}, \citenamefont {Kurlov}, \citenamefont {Gritsev},
  \citenamefont {Fedorov},\ and\ \citenamefont {Krimer}}]{Bakker2022}%
  \BibitemOpen
  \bibfield  {author} {\bibinfo {author} {\bibfnamefont {L.~R.}\ \bibnamefont
  {Bakker}}, \bibinfo {author} {\bibfnamefont {M.~S.}\ \bibnamefont
  {Bahovadinov}}, \bibinfo {author} {\bibfnamefont {D.~V.}\ \bibnamefont
  {Kurlov}}, \bibinfo {author} {\bibfnamefont {V.}~\bibnamefont {Gritsev}},
  \bibinfo {author} {\bibfnamefont {A.~K.}\ \bibnamefont {Fedorov}},\ and\
  \bibinfo {author} {\bibfnamefont {D.~O.}\ \bibnamefont {Krimer}},\ }\bibfield
   {title} {\bibinfo {title} {{Driven-Dissipative Time Crystalline Phases in a
  Two-Mode Bosonic System with Kerr Nonlinearity}},\ }\href
  {https://doi.org/10.1103/PhysRevLett.129.250401} {\bibfield  {journal}
  {\bibinfo  {journal} {Phys. Rev. Lett.}\ }\textbf {\bibinfo {volume} {129}},\
  \bibinfo {pages} {250401} (\bibinfo {year} {2022})}\BibitemShut {NoStop}%
\bibitem [{\citenamefont {Krishna}\ \emph {et~al.}(2023)\citenamefont
  {Krishna}, \citenamefont {Solanki}, \citenamefont
  {Hajdu\ifmmode~\check{s}\else \v{s}\fi{}ek},\ and\ \citenamefont
  {Vinjanampathy}}]{Krishna2023}%
  \BibitemOpen
  \bibfield  {author} {\bibinfo {author} {\bibfnamefont {M.}~\bibnamefont
  {Krishna}}, \bibinfo {author} {\bibfnamefont {P.}~\bibnamefont {Solanki}},
  \bibinfo {author} {\bibfnamefont {M.}~\bibnamefont
  {Hajdu\ifmmode~\check{s}\else \v{s}\fi{}ek}},\ and\ \bibinfo {author}
  {\bibfnamefont {S.}~\bibnamefont {Vinjanampathy}},\ }\bibfield  {title}
  {\bibinfo {title} {{Measurement-Induced Continuous Time Crystals}},\ }\href
  {https://doi.org/10.1103/PhysRevLett.130.150401} {\bibfield  {journal}
  {\bibinfo  {journal} {Phys. Rev. Lett.}\ }\textbf {\bibinfo {volume} {130}},\
  \bibinfo {pages} {150401} (\bibinfo {year} {2023})}\BibitemShut {NoStop}%
\bibitem [{\citenamefont {Kongkhambut}\ \emph {et~al.}(2022)\citenamefont
  {Kongkhambut}, \citenamefont {Skulte}, \citenamefont {Mathey}, \citenamefont
  {Cosme}, \citenamefont {Hemmerich},\ and\ \citenamefont
  {Ke{\ss}ler}}]{Kongkhambut2022}%
  \BibitemOpen
  \bibfield  {author} {\bibinfo {author} {\bibfnamefont {P.}~\bibnamefont
  {Kongkhambut}}, \bibinfo {author} {\bibfnamefont {J.}~\bibnamefont {Skulte}},
  \bibinfo {author} {\bibfnamefont {L.}~\bibnamefont {Mathey}}, \bibinfo
  {author} {\bibfnamefont {J.~G.}\ \bibnamefont {Cosme}}, \bibinfo {author}
  {\bibfnamefont {A.}~\bibnamefont {Hemmerich}},\ and\ \bibinfo {author}
  {\bibfnamefont {H.}~\bibnamefont {Ke{\ss}ler}},\ }\bibfield  {title}
  {\bibinfo {title} {{Observation of a continuous time crystal}},\ }\href
  {https://doi.org/10.1126/science.abo3382} {\bibfield  {journal} {\bibinfo
  {journal} {Science}\ }\textbf {\bibinfo {volume} {377}},\ \bibinfo {pages}
  {670} (\bibinfo {year} {2022})}\BibitemShut {NoStop}%
\bibitem [{\citenamefont {Liu}\ \emph {et~al.}(2023)\citenamefont {Liu},
  \citenamefont {Ou}, \citenamefont {MacDonald},\ and\ \citenamefont
  {Zheludev}}]{Liu2023}%
  \BibitemOpen
  \bibfield  {author} {\bibinfo {author} {\bibfnamefont {T.}~\bibnamefont
  {Liu}}, \bibinfo {author} {\bibfnamefont {J.-Y.}\ \bibnamefont {Ou}},
  \bibinfo {author} {\bibfnamefont {K.}~\bibnamefont {MacDonald}},\ and\
  \bibinfo {author} {\bibfnamefont {N.}~\bibnamefont {Zheludev}},\ }\bibfield
  {title} {\bibinfo {title} {{Photonic metamaterial analogue of a continuous
  time crystal}},\ }\bibfield  {journal} {\bibinfo  {journal} {Nature Physics}\
  }\href {https://doi.org/10.1038/s41567-023-02023-5}
  {10.1038/s41567-023-02023-5} (\bibinfo {year} {2023})\BibitemShut {NoStop}%
\bibitem [{\citenamefont {Chen}\ and\ \citenamefont {Zhang}(2023)}]{Chen2023}%
  \BibitemOpen
  \bibfield  {author} {\bibinfo {author} {\bibfnamefont {Y.-H.}\ \bibnamefont
  {Chen}}\ and\ \bibinfo {author} {\bibfnamefont {X.}~\bibnamefont {Zhang}},\
  }\bibfield  {title} {\bibinfo {title} {Realization of an inherent time
  crystal in a dissipative many-body system},\ }\href@noop {} {\bibfield
  {journal} {\bibinfo  {journal} {Nature Communications}\ }\textbf {\bibinfo
  {volume} {14}},\ \bibinfo {pages} {6161} (\bibinfo {year}
  {2023})}\BibitemShut {NoStop}%
\bibitem [{\citenamefont {Zaletel}\ \emph {et~al.}(2023)\citenamefont
  {Zaletel}, \citenamefont {Lukin}, \citenamefont {Monroe}, \citenamefont
  {Nayak}, \citenamefont {Wilczek},\ and\ \citenamefont {Yao}}]{zaletel2023}%
  \BibitemOpen
  \bibfield  {author} {\bibinfo {author} {\bibfnamefont {M.~P.}\ \bibnamefont
  {Zaletel}}, \bibinfo {author} {\bibfnamefont {M.}~\bibnamefont {Lukin}},
  \bibinfo {author} {\bibfnamefont {C.}~\bibnamefont {Monroe}}, \bibinfo
  {author} {\bibfnamefont {C.}~\bibnamefont {Nayak}}, \bibinfo {author}
  {\bibfnamefont {F.}~\bibnamefont {Wilczek}},\ and\ \bibinfo {author}
  {\bibfnamefont {N.~Y.}\ \bibnamefont {Yao}},\ }\bibfield  {title} {\bibinfo
  {title} {{Colloquium: Quantum and classical discrete time crystals}},\ }\href
  {https://doi.org/10.1103/RevModPhys.95.031001} {\bibfield  {journal}
  {\bibinfo  {journal} {Rev. Mod. Phys.}\ }\textbf {\bibinfo {volume} {95}},\
  \bibinfo {pages} {031001} (\bibinfo {year} {2023})}\BibitemShut {NoStop}%
\bibitem [{\citenamefont {Keeling}\ \emph {et~al.}(2010)\citenamefont
  {Keeling}, \citenamefont {Bhaseen},\ and\ \citenamefont
  {Simons}}]{Keeling2010}%
  \BibitemOpen
  \bibfield  {author} {\bibinfo {author} {\bibfnamefont {J.}~\bibnamefont
  {Keeling}}, \bibinfo {author} {\bibfnamefont {M.}~\bibnamefont {Bhaseen}},\
  and\ \bibinfo {author} {\bibfnamefont {B.}~\bibnamefont {Simons}},\
  }\bibfield  {title} {\bibinfo {title} {{Collective Dynamics of Bose-Einstein
  Condensates in Optical Cavities}},\ }\href
  {https://doi.org/10.1103/PhysRevLett.105.043001} {\bibfield  {journal}
  {\bibinfo  {journal} {Phys. Rev. Lett.}\ }\textbf {\bibinfo {volume} {105}},\
  \bibinfo {pages} {043001} (\bibinfo {year} {2010})}\BibitemShut {NoStop}%
\bibitem [{\citenamefont {Bhaseen}\ \emph {et~al.}(2012)\citenamefont
  {Bhaseen}, \citenamefont {Mayoh}, \citenamefont {Simons},\ and\ \citenamefont
  {Keeling}}]{Bhaseen2012}%
  \BibitemOpen
  \bibfield  {author} {\bibinfo {author} {\bibfnamefont {M.}~\bibnamefont
  {Bhaseen}}, \bibinfo {author} {\bibfnamefont {J.}~\bibnamefont {Mayoh}},
  \bibinfo {author} {\bibfnamefont {B.~D.}\ \bibnamefont {Simons}},\ and\
  \bibinfo {author} {\bibfnamefont {J.}~\bibnamefont {Keeling}},\ }\bibfield
  {title} {\bibinfo {title} {{Dynamics of nonequilibrium Dicke models}},\
  }\href {https://doi.org/10.1103/PhysRevA.85.013817} {\bibfield  {journal}
  {\bibinfo  {journal} {Phys. Rev. A}\ }\textbf {\bibinfo {volume} {85}},\
  \bibinfo {pages} {013817} (\bibinfo {year} {2012})}\BibitemShut {NoStop}%
\bibitem [{\citenamefont {Piazza}\ and\ \citenamefont
  {Ritsch}(2015)}]{Piazza2015}%
  \BibitemOpen
  \bibfield  {author} {\bibinfo {author} {\bibfnamefont {F.}~\bibnamefont
  {Piazza}}\ and\ \bibinfo {author} {\bibfnamefont {H.}~\bibnamefont
  {Ritsch}},\ }\bibfield  {title} {\bibinfo {title} {{Self-Ordered Limit
  Cycles, Chaos, and Phase Slippage with a Superfluid inside an Optical
  Resonator}},\ }\href {https://doi.org/10.1103/PhysRevLett.115.163601}
  {\bibfield  {journal} {\bibinfo  {journal} {Phys. Rev. Lett.}\ }\textbf
  {\bibinfo {volume} {115}},\ \bibinfo {pages} {163601} (\bibinfo {year}
  {2015})}\BibitemShut {NoStop}%
\bibitem [{\citenamefont {Zare~Rameshti}\ \emph {et~al.}(2022)\citenamefont
  {Zare~Rameshti}, \citenamefont {Kusminskiy}, \citenamefont {Haigh},
  \citenamefont {Usami}, \citenamefont {Lachance-Quirion}, \citenamefont
  {Nakamura}, \citenamefont {Hu}, \citenamefont {Tang}, \citenamefont {Bauer},\
  and\ \citenamefont {Blanter}}]{Zare2022}%
  \BibitemOpen
  \bibfield  {author} {\bibinfo {author} {\bibfnamefont {B.}~\bibnamefont
  {Zare~Rameshti}}, \bibinfo {author} {\bibfnamefont {S.}~\bibnamefont
  {Kusminskiy}}, \bibinfo {author} {\bibfnamefont {J.}~\bibnamefont {Haigh}},
  \bibinfo {author} {\bibfnamefont {K.}~\bibnamefont {Usami}}, \bibinfo
  {author} {\bibfnamefont {D.}~\bibnamefont {Lachance-Quirion}}, \bibinfo
  {author} {\bibfnamefont {Y.}~\bibnamefont {Nakamura}}, \bibinfo {author}
  {\bibfnamefont {C.-M.}\ \bibnamefont {Hu}}, \bibinfo {author} {\bibfnamefont
  {H.}~\bibnamefont {Tang}}, \bibinfo {author} {\bibfnamefont {G.}~\bibnamefont
  {Bauer}},\ and\ \bibinfo {author} {\bibfnamefont {Y.}~\bibnamefont
  {Blanter}},\ }\bibfield  {title} {\bibinfo {title} {{Cavity magnonics}},\
  }\href {https://doi.org/10.1016/j.physrep.2022.06.001} {\bibfield  {journal}
  {\bibinfo  {journal} {Physics Reports}\ }\textbf {\bibinfo {volume} {979}},\
  \bibinfo {pages} {1} (\bibinfo {year} {2022})}\BibitemShut {NoStop}%
\bibitem [{\citenamefont {Chang}\ \emph {et~al.}(2020)\citenamefont {Chang},
  \citenamefont {Sab\'{\i}n}, \citenamefont {Forn-D\'{\i}az}, \citenamefont
  {Quijandr\'{\i}a}, \citenamefont {Vadiraj}, \citenamefont {Nsanzineza},
  \citenamefont {Johansson},\ and\ \citenamefont {Wilson}}]{Chang2020}%
  \BibitemOpen
  \bibfield  {author} {\bibinfo {author} {\bibfnamefont {C.~W.}\ \bibnamefont
  {Chang}}, \bibinfo {author} {\bibfnamefont {C.}~\bibnamefont {Sab\'{\i}n}},
  \bibinfo {author} {\bibfnamefont {P.}~\bibnamefont {Forn-D\'{\i}az}},
  \bibinfo {author} {\bibfnamefont {F.}~\bibnamefont {Quijandr\'{\i}a}},
  \bibinfo {author} {\bibfnamefont {A.~M.}\ \bibnamefont {Vadiraj}}, \bibinfo
  {author} {\bibfnamefont {I.}~\bibnamefont {Nsanzineza}}, \bibinfo {author}
  {\bibfnamefont {G.}~\bibnamefont {Johansson}},\ and\ \bibinfo {author}
  {\bibfnamefont {C.}~\bibnamefont {Wilson}},\ }\bibfield  {title} {\bibinfo
  {title} {{Observation of Three-Photon Spontaneous Parametric Down-Conversion
  in a Superconducting Parametric Cavity}},\ }\href
  {https://doi.org/10.1103/PhysRevX.10.011011} {\bibfield  {journal} {\bibinfo
  {journal} {Phys. Rev. X}\ }\textbf {\bibinfo {volume} {10}},\ \bibinfo
  {pages} {011011} (\bibinfo {year} {2020})}\BibitemShut {NoStop}%
\bibitem [{\citenamefont {Minganti}\ \emph {et~al.}(2023)\citenamefont
  {Minganti}, \citenamefont {Garbe}, \citenamefont {Le~Boit\'e},\ and\
  \citenamefont {Felicetti}}]{Minganti2023}%
  \BibitemOpen
  \bibfield  {author} {\bibinfo {author} {\bibfnamefont {F.}~\bibnamefont
  {Minganti}}, \bibinfo {author} {\bibfnamefont {L.}~\bibnamefont {Garbe}},
  \bibinfo {author} {\bibfnamefont {A.}~\bibnamefont {Le~Boit\'e}},\ and\
  \bibinfo {author} {\bibfnamefont {S.}~\bibnamefont {Felicetti}},\ }\bibfield
  {title} {\bibinfo {title} {{Non-Gaussian superradiant transition via
  three-body ultrastrong coupling}},\ }\href
  {https://doi.org/10.1103/PhysRevA.107.013715} {\bibfield  {journal} {\bibinfo
   {journal} {Phys. Rev. A}\ }\textbf {\bibinfo {volume} {107}},\ \bibinfo
  {pages} {013715} (\bibinfo {year} {2023})}\BibitemShut {NoStop}%
\bibitem [{\citenamefont {Larson}\ and\ \citenamefont
  {Mavrogordatos}(2021)}]{larson2021}%
  \BibitemOpen
  \bibfield  {author} {\bibinfo {author} {\bibfnamefont {J.}~\bibnamefont
  {Larson}}\ and\ \bibinfo {author} {\bibfnamefont {T.}~\bibnamefont
  {Mavrogordatos}},\ }\href
  {https://iopscience.iop.org/book/mono/978-0-7503-3447-1} {\emph {\bibinfo
  {title} {The Jaynes–Cummings Model and Its Descendants}}}\ (\bibinfo
  {publisher} {IOP Publishing Ltd},\ \bibinfo {year} {2021})\BibitemShut
  {NoStop}%
\bibitem [{\citenamefont {Aspelmeyer}\ \emph {et~al.}(2014)\citenamefont
  {Aspelmeyer}, \citenamefont {Kippenberg},\ and\ \citenamefont
  {Marquardt}}]{aspelmeyer2014}%
  \BibitemOpen
  \bibfield  {author} {\bibinfo {author} {\bibfnamefont {M.}~\bibnamefont
  {Aspelmeyer}}, \bibinfo {author} {\bibfnamefont {T.~J.}\ \bibnamefont
  {Kippenberg}},\ and\ \bibinfo {author} {\bibfnamefont {F.}~\bibnamefont
  {Marquardt}},\ }\bibfield  {title} {\bibinfo {title} {Cavity optomechanics},\
  }\href {https://doi.org/10.1103/RevModPhys.86.1391} {\bibfield  {journal}
  {\bibinfo  {journal} {Reviews of Modern Physics}\ }\textbf {\bibinfo {volume}
  {86}},\ \bibinfo {pages} {1391} (\bibinfo {year} {2014})}\BibitemShut
  {NoStop}%
\bibitem [{\citenamefont {Rameshti}\ \emph {et~al.}(2022)\citenamefont
  {Rameshti}, \citenamefont {Kusminskiy}, \citenamefont {Haigh}, \citenamefont
  {Usami}, \citenamefont {Lachance-Quirion}, \citenamefont {Nakamura},
  \citenamefont {Hu}, \citenamefont {Tang}, \citenamefont {Bauer},\ and\
  \citenamefont {Blanter}}]{Rameshti2022}%
  \BibitemOpen
  \bibfield  {author} {\bibinfo {author} {\bibfnamefont {B.~Z.}\ \bibnamefont
  {Rameshti}}, \bibinfo {author} {\bibfnamefont {S.~V.}\ \bibnamefont
  {Kusminskiy}}, \bibinfo {author} {\bibfnamefont {J.~A.}\ \bibnamefont
  {Haigh}}, \bibinfo {author} {\bibfnamefont {K.}~\bibnamefont {Usami}},
  \bibinfo {author} {\bibfnamefont {D.}~\bibnamefont {Lachance-Quirion}},
  \bibinfo {author} {\bibfnamefont {Y.}~\bibnamefont {Nakamura}}, \bibinfo
  {author} {\bibfnamefont {C.~M.}\ \bibnamefont {Hu}}, \bibinfo {author}
  {\bibfnamefont {H.~X.}\ \bibnamefont {Tang}}, \bibinfo {author}
  {\bibfnamefont {G.~E.}\ \bibnamefont {Bauer}},\ and\ \bibinfo {author}
  {\bibfnamefont {Y.~M.}\ \bibnamefont {Blanter}},\ }\href
  {https://doi.org/10.1016/j.physrep.2022.06.001} {\bibinfo {title} {Cavity
  magnonics}} (\bibinfo {year} {2022})\BibitemShut {NoStop}%
\bibitem [{\citenamefont {Kosior}\ \emph {et~al.}(2022)\citenamefont {Kosior},
  \citenamefont {Ritsch},\ and\ \citenamefont {Mivehvar}}]{kosior2022}%
  \BibitemOpen
  \bibfield  {author} {\bibinfo {author} {\bibfnamefont {A.}~\bibnamefont
  {Kosior}}, \bibinfo {author} {\bibfnamefont {H.}~\bibnamefont {Ritsch}},\
  and\ \bibinfo {author} {\bibfnamefont {F.}~\bibnamefont {Mivehvar}},\
  }\href@noop {} {\bibinfo {title} {{Nonequilibrium phases of ultracold bosons
  with cavity-induced dynamic gauge fields}}} (\bibinfo {year} {2022}),\
  \Eprint {https://arxiv.org/abs/2208.04602} {arXiv:2208.04602
  [cond-mat.quant-gas]} \BibitemShut {NoStop}%
\bibitem [{\citenamefont {Strogatz}(2000)}]{Strogatz2000}%
  \BibitemOpen
  \bibfield  {author} {\bibinfo {author} {\bibfnamefont {S.}~\bibnamefont
  {Strogatz}},\ }\href {https://books.google.de/books?id=NZZDnQEACAAJ} {\emph
  {\bibinfo {title} {{Nonlinear Dynamics and Chaos: With Applications to
  Physics, Biology, Chemistry and Engineering}}}},\ Studies in nonlinearity\
  (\bibinfo  {publisher} {Westview},\ \bibinfo {year} {2000})\BibitemShut
  {NoStop}%
\bibitem [{\citenamefont {Dreon}\ \emph {et~al.}(2022)\citenamefont {Dreon},
  \citenamefont {Baumg{\"a}rtner}, \citenamefont {Hertlein}, \citenamefont
  {Esslinger},\ and\ \citenamefont {Donner}}]{Dreon2022}%
  \BibitemOpen
  \bibfield  {author} {\bibinfo {author} {\bibfnamefont {D.}~\bibnamefont
  {Dreon}}, \bibinfo {author} {\bibfnamefont {X.}~\bibnamefont
  {Baumg{\"a}rtner}, \bibfnamefont {A.and~Li}}, \bibinfo {author}
  {\bibfnamefont {S.}~\bibnamefont {Hertlein}}, \bibinfo {author}
  {\bibfnamefont {T.}~\bibnamefont {Esslinger}},\ and\ \bibinfo {author}
  {\bibfnamefont {T.}~\bibnamefont {Donner}},\ }\bibfield  {title} {\bibinfo
  {title} {{Self-oscillating pump in a topological dissipative atom-cavity
  system}},\ }\href {https://doi.org/10.1038/s41586-022-04970-0} {\bibfield
  {journal} {\bibinfo  {journal} {Nature}\ }\textbf {\bibinfo {volume} {608}},\
  \bibinfo {pages} {494} (\bibinfo {year} {2022})}\BibitemShut {NoStop}%
\bibitem [{\citenamefont {Nie}\ and\ \citenamefont {Zheng}(2023)}]{Nie2023}%
  \BibitemOpen
  \bibfield  {author} {\bibinfo {author} {\bibfnamefont {X.}~\bibnamefont
  {Nie}}\ and\ \bibinfo {author} {\bibfnamefont {W.}~\bibnamefont {Zheng}},\
  }\bibfield  {title} {\bibinfo {title} {Nonequilibrium phases of a fermi gas
  inside a cavity with imbalanced pumping},\ }\href
  {https://doi.org/10.1103/PhysRevA.108.043312} {\bibfield  {journal} {\bibinfo
   {journal} {Phys. Rev. A}\ }\textbf {\bibinfo {volume} {108}},\ \bibinfo
  {pages} {043312} (\bibinfo {year} {2023})}\BibitemShut {NoStop}%
\bibitem [{\citenamefont {Baumann}\ \emph {et~al.}(2010)\citenamefont
  {Baumann}, \citenamefont {Guerlin}, \citenamefont {Brennecke},\ and\
  \citenamefont {Esslinger}}]{Baumann2010}%
  \BibitemOpen
  \bibfield  {author} {\bibinfo {author} {\bibfnamefont {K.}~\bibnamefont
  {Baumann}}, \bibinfo {author} {\bibfnamefont {C.}~\bibnamefont {Guerlin}},
  \bibinfo {author} {\bibfnamefont {F.}~\bibnamefont {Brennecke}},\ and\
  \bibinfo {author} {\bibfnamefont {T.}~\bibnamefont {Esslinger}},\ }\bibfield
  {title} {\bibinfo {title} {Dicke quantum phase transition with a superfluid
  gas in an optical cavity},\ }\href {https://doi.org/10.1038/nature09009}
  {\bibfield  {journal} {\bibinfo  {journal} {Nature}\ }\textbf {\bibinfo
  {volume} {464}},\ \bibinfo {pages} {1301} (\bibinfo {year}
  {2010})}\BibitemShut {NoStop}%
\bibitem [{\citenamefont {Klinder}\ \emph {et~al.}(2015)\citenamefont
  {Klinder}, \citenamefont {Ke{\ss}ler}, \citenamefont {Wolke}, \citenamefont
  {Mathey},\ and\ \citenamefont {Hemmerich}}]{Klinder2015}%
  \BibitemOpen
  \bibfield  {author} {\bibinfo {author} {\bibfnamefont {J.}~\bibnamefont
  {Klinder}}, \bibinfo {author} {\bibfnamefont {H.}~\bibnamefont {Ke{\ss}ler}},
  \bibinfo {author} {\bibfnamefont {M.}~\bibnamefont {Wolke}}, \bibinfo
  {author} {\bibfnamefont {L.}~\bibnamefont {Mathey}},\ and\ \bibinfo {author}
  {\bibfnamefont {A.}~\bibnamefont {Hemmerich}},\ }\bibfield  {title} {\bibinfo
  {title} {{Dynamical phase transition in the open Dicke model}},\ }\href
  {https://doi.org/10.1073/pnas.1417132112} {\bibfield  {journal} {\bibinfo
  {journal} {Proceedings of the National Academy of Sciences}\ }\textbf
  {\bibinfo {volume} {112}},\ \bibinfo {pages} {3290} (\bibinfo {year}
  {2015})}\BibitemShut {NoStop}%
\bibitem [{\citenamefont {Gao}\ \emph {et~al.}(2023)\citenamefont {Gao},
  \citenamefont {Zhou}, \citenamefont {Guo},\ and\ \citenamefont
  {Luo}}]{Gao2023}%
  \BibitemOpen
  \bibfield  {author} {\bibinfo {author} {\bibfnamefont {P.}~\bibnamefont
  {Gao}}, \bibinfo {author} {\bibfnamefont {Z.-W.}\ \bibnamefont {Zhou}},
  \bibinfo {author} {\bibfnamefont {G.-C.}\ \bibnamefont {Guo}},\ and\ \bibinfo
  {author} {\bibfnamefont {X.-W.}\ \bibnamefont {Luo}},\ }\bibfield  {title}
  {\bibinfo {title} {{Self-organized limit cycles in red-detuned atom-cavity
  systems}},\ }\href {https://doi.org/10.1103/PhysRevA.107.023311} {\bibfield
  {journal} {\bibinfo  {journal} {Phys. Rev. A}\ }\textbf {\bibinfo {volume}
  {107}},\ \bibinfo {pages} {023311} (\bibinfo {year} {2023})}\BibitemShut
  {NoStop}%
\bibitem [{\citenamefont {Ke{\ss}ler}\ \emph {et~al.}(2014)\citenamefont
  {Ke{\ss}ler}, \citenamefont {Klinder}, \citenamefont {Wolke},\ and\
  \citenamefont {Hemmerich}}]{Kessler2014}%
  \BibitemOpen
  \bibfield  {author} {\bibinfo {author} {\bibfnamefont {H.}~\bibnamefont
  {Ke{\ss}ler}}, \bibinfo {author} {\bibfnamefont {J.}~\bibnamefont {Klinder}},
  \bibinfo {author} {\bibfnamefont {M.}~\bibnamefont {Wolke}},\ and\ \bibinfo
  {author} {\bibfnamefont {A.}~\bibnamefont {Hemmerich}},\ }\bibfield  {title}
  {\bibinfo {title} {Optomechanical atom-cavity interaction in the sub-recoil
  regime},\ }\href {https://doi.org/10.1088/1367-2630/16/5/053008} {\bibfield
  {journal} {\bibinfo  {journal} {New Journal of Physics}\ }\textbf {\bibinfo
  {volume} {16}},\ \bibinfo {pages} {053008} (\bibinfo {year}
  {2014})}\BibitemShut {NoStop}%
\bibitem [{\citenamefont {Klinder}\ \emph {et~al.}(2016)\citenamefont
  {Klinder}, \citenamefont {Ke{\ss}ler}, \citenamefont {Georges}, \citenamefont
  {Vargas},\ and\ \citenamefont {Hemmerich}}]{Klinder2016}%
  \BibitemOpen
  \bibfield  {author} {\bibinfo {author} {\bibfnamefont {J.}~\bibnamefont
  {Klinder}}, \bibinfo {author} {\bibfnamefont {H.}~\bibnamefont {Ke{\ss}ler}},
  \bibinfo {author} {\bibfnamefont {C.}~\bibnamefont {Georges}}, \bibinfo
  {author} {\bibfnamefont {J.}~\bibnamefont {Vargas}},\ and\ \bibinfo {author}
  {\bibfnamefont {A.}~\bibnamefont {Hemmerich}},\ }\bibfield  {title} {\bibinfo
  {title} {Bose--einstein condensates in an optical cavity with sub-recoil
  bandwidth},\ }\href@noop {} {\bibfield  {journal} {\bibinfo  {journal}
  {Applied Physics B}\ }\textbf {\bibinfo {volume} {122}},\ \bibinfo {pages}
  {299} (\bibinfo {year} {2016})}\BibitemShut {NoStop}%
\bibitem [{\citenamefont {Ritsch}\ \emph {et~al.}(2013)\citenamefont {Ritsch},
  \citenamefont {Domokos}, \citenamefont {Brennecke},\ and\ \citenamefont
  {Esslinger}}]{Ritsch2013}%
  \BibitemOpen
  \bibfield  {author} {\bibinfo {author} {\bibfnamefont {H.}~\bibnamefont
  {Ritsch}}, \bibinfo {author} {\bibfnamefont {P.}~\bibnamefont {Domokos}},
  \bibinfo {author} {\bibfnamefont {F.}~\bibnamefont {Brennecke}},\ and\
  \bibinfo {author} {\bibfnamefont {T.}~\bibnamefont {Esslinger}},\ }\bibfield
  {title} {\bibinfo {title} {{Cold atoms in cavity-generated dynamical optical
  potentials}},\ }\href {https://doi.org/10.1103/RevModPhys.85.553} {\bibfield
  {journal} {\bibinfo  {journal} {Rev. Mod. Phys.}\ }\textbf {\bibinfo {volume}
  {85}},\ \bibinfo {pages} {553} (\bibinfo {year} {2013})}\BibitemShut
  {NoStop}%
\bibitem [{\citenamefont {Cosme}\ \emph {et~al.}(2019)\citenamefont {Cosme},
  \citenamefont {Skulte},\ and\ \citenamefont {Mathey}}]{Cosme2019}%
  \BibitemOpen
  \bibfield  {author} {\bibinfo {author} {\bibfnamefont {J.~G.}\ \bibnamefont
  {Cosme}}, \bibinfo {author} {\bibfnamefont {J.}~\bibnamefont {Skulte}},\ and\
  \bibinfo {author} {\bibfnamefont {L.}~\bibnamefont {Mathey}},\ }\bibfield
  {title} {\bibinfo {title} {{Time crystals in a shaken atom-cavity system}},\
  }\href {https://doi.org/10.1103/PhysRevA.100.053615} {\bibfield  {journal}
  {\bibinfo  {journal} {Phys. Rev. A}\ }\textbf {\bibinfo {volume} {100}},\
  \bibinfo {pages} {053615} (\bibinfo {year} {2019})}\BibitemShut {NoStop}%
\bibitem [{\citenamefont {Tuquero}\ \emph {et~al.}(2022)\citenamefont
  {Tuquero}, \citenamefont {Skulte}, \citenamefont {Mathey},\ and\
  \citenamefont {Cosme}}]{Tuquero2022}%
  \BibitemOpen
  \bibfield  {author} {\bibinfo {author} {\bibfnamefont {R.~J.~L.}\
  \bibnamefont {Tuquero}}, \bibinfo {author} {\bibfnamefont {J.}~\bibnamefont
  {Skulte}}, \bibinfo {author} {\bibfnamefont {L.}~\bibnamefont {Mathey}},\
  and\ \bibinfo {author} {\bibfnamefont {J.~G.}\ \bibnamefont {Cosme}},\
  }\bibfield  {title} {\bibinfo {title} {{Dissipative time crystal in an
  atom-cavity system: Influence of trap and competing interactions}},\ }\href
  {https://doi.org/10.1103/PhysRevA.105.043311} {\bibfield  {journal} {\bibinfo
   {journal} {Phys. Rev. A}\ }\textbf {\bibinfo {volume} {105}},\ \bibinfo
  {pages} {043311} (\bibinfo {year} {2022})}\BibitemShut {NoStop}%
\bibitem [{\citenamefont {Wolf}\ \emph {et~al.}(2018)\citenamefont {Wolf},
  \citenamefont {Schuster}, \citenamefont {Schmidt}, \citenamefont {Slama},\
  and\ \citenamefont {Zimmermann}}]{Wolf2018}%
  \BibitemOpen
  \bibfield  {author} {\bibinfo {author} {\bibfnamefont {P.}~\bibnamefont
  {Wolf}}, \bibinfo {author} {\bibfnamefont {S.~C.}\ \bibnamefont {Schuster}},
  \bibinfo {author} {\bibfnamefont {D.}~\bibnamefont {Schmidt}}, \bibinfo
  {author} {\bibfnamefont {S.}~\bibnamefont {Slama}},\ and\ \bibinfo {author}
  {\bibfnamefont {C.}~\bibnamefont {Zimmermann}},\ }\bibfield  {title}
  {\bibinfo {title} {Observation of subradiant atomic momentum states with
  bose-einstein condensates in a recoil resolving optical ring resonator},\
  }\href {https://doi.org/10.1103/PhysRevLett.121.173602} {\bibfield  {journal}
  {\bibinfo  {journal} {Phys. Rev. Lett.}\ }\textbf {\bibinfo {volume} {121}},\
  \bibinfo {pages} {173602} (\bibinfo {year} {2018})}\BibitemShut {NoStop}%
\bibitem [{\citenamefont {Skulte}\ \emph {et~al.}(2021)\citenamefont {Skulte},
  \citenamefont {Kongkhambut}, \citenamefont {Ke\ss{}ler}, \citenamefont
  {Hemmerich}, \citenamefont {Mathey},\ and\ \citenamefont
  {Cosme}}]{Skulte2021}%
  \BibitemOpen
  \bibfield  {author} {\bibinfo {author} {\bibfnamefont {J.}~\bibnamefont
  {Skulte}}, \bibinfo {author} {\bibfnamefont {P.}~\bibnamefont {Kongkhambut}},
  \bibinfo {author} {\bibfnamefont {H.}~\bibnamefont {Ke\ss{}ler}}, \bibinfo
  {author} {\bibfnamefont {A.}~\bibnamefont {Hemmerich}}, \bibinfo {author}
  {\bibfnamefont {L.}~\bibnamefont {Mathey}},\ and\ \bibinfo {author}
  {\bibfnamefont {J.~G.}\ \bibnamefont {Cosme}},\ }\bibfield  {title} {\bibinfo
  {title} {{Parametrically driven dissipative three-level Dicke model}},\
  }\href {https://doi.org/10.1103/PhysRevA.104.063705} {\bibfield  {journal}
  {\bibinfo  {journal} {Phys. Rev. A}\ }\textbf {\bibinfo {volume} {104}},\
  \bibinfo {pages} {063705} (\bibinfo {year} {2021})}\BibitemShut {NoStop}%
\bibitem [{\citenamefont {Kongkhambut}\ \emph {et~al.}(2021)\citenamefont
  {Kongkhambut}, \citenamefont {Ke\ss{}ler}, \citenamefont {Skulte},
  \citenamefont {Mathey}, \citenamefont {Cosme},\ and\ \citenamefont
  {Hemmerich}}]{Kongkhambut2021}%
  \BibitemOpen
  \bibfield  {author} {\bibinfo {author} {\bibfnamefont {P.}~\bibnamefont
  {Kongkhambut}}, \bibinfo {author} {\bibfnamefont {H.}~\bibnamefont
  {Ke\ss{}ler}}, \bibinfo {author} {\bibfnamefont {J.}~\bibnamefont {Skulte}},
  \bibinfo {author} {\bibfnamefont {L.}~\bibnamefont {Mathey}}, \bibinfo
  {author} {\bibfnamefont {J.~G.}\ \bibnamefont {Cosme}},\ and\ \bibinfo
  {author} {\bibfnamefont {A.}~\bibnamefont {Hemmerich}},\ }\bibfield  {title}
  {\bibinfo {title} {{Realization of a Periodically Driven Open Three-Level
  Dicke Model}},\ }\href {https://doi.org/10.1103/PhysRevLett.127.253601}
  {\bibfield  {journal} {\bibinfo  {journal} {Phys. Rev. Lett.}\ }\textbf
  {\bibinfo {volume} {127}},\ \bibinfo {pages} {253601} (\bibinfo {year}
  {2021})}\BibitemShut {NoStop}%
\bibitem [{\citenamefont {Lin}\ \emph {et~al.}(2022)\citenamefont {Lin},
  \citenamefont {Rosa-Medina}, \citenamefont {Ferri}, \citenamefont {Finger},
  \citenamefont {Kroeger}, \citenamefont {Donner}, \citenamefont {Esslinger},\
  and\ \citenamefont {Chitra}}]{Lin2022}%
  \BibitemOpen
  \bibfield  {author} {\bibinfo {author} {\bibfnamefont {R.}~\bibnamefont
  {Lin}}, \bibinfo {author} {\bibfnamefont {R.}~\bibnamefont {Rosa-Medina}},
  \bibinfo {author} {\bibfnamefont {F.}~\bibnamefont {Ferri}}, \bibinfo
  {author} {\bibfnamefont {F.}~\bibnamefont {Finger}}, \bibinfo {author}
  {\bibfnamefont {K.}~\bibnamefont {Kroeger}}, \bibinfo {author} {\bibfnamefont
  {T.}~\bibnamefont {Donner}}, \bibinfo {author} {\bibfnamefont
  {T.}~\bibnamefont {Esslinger}},\ and\ \bibinfo {author} {\bibfnamefont
  {R.}~\bibnamefont {Chitra}},\ }\bibfield  {title} {\bibinfo {title}
  {{Dissipation-Engineered Family of Nearly Dark States in Many-Body
  Cavity-Atom Systems}},\ }\href
  {https://doi.org/10.1103/PhysRevLett.128.153601} {\bibfield  {journal}
  {\bibinfo  {journal} {Phys. Rev. Lett.}\ }\textbf {\bibinfo {volume} {128}},\
  \bibinfo {pages} {153601} (\bibinfo {year} {2022})}\BibitemShut {NoStop}%
\bibitem [{\citenamefont {Fan}\ and\ \citenamefont {Jia}(2023)}]{Fan2023}%
  \BibitemOpen
  \bibfield  {author} {\bibinfo {author} {\bibfnamefont {J.}~\bibnamefont
  {Fan}}\ and\ \bibinfo {author} {\bibfnamefont {S.}~\bibnamefont {Jia}},\
  }\bibfield  {title} {\bibinfo {title} {{Collective dynamics of the unbalanced
  three-level Dicke model}},\ }\href
  {https://doi.org/10.1103/PhysRevA.107.033711} {\bibfield  {journal} {\bibinfo
   {journal} {Phys. Rev. A}\ }\textbf {\bibinfo {volume} {107}},\ \bibinfo
  {pages} {033711} (\bibinfo {year} {2023})}\BibitemShut {NoStop}%
\bibitem [{\citenamefont {Skulte}\ \emph {et~al.}(2023)\citenamefont {Skulte},
  \citenamefont {Kongkhambut}, \citenamefont {Rao}, \citenamefont {Mathey},
  \citenamefont {Ke\ss{}ler}, \citenamefont {Hemmerich},\ and\ \citenamefont
  {Cosme}}]{Skulte2023}%
  \BibitemOpen
  \bibfield  {author} {\bibinfo {author} {\bibfnamefont {J.}~\bibnamefont
  {Skulte}}, \bibinfo {author} {\bibfnamefont {P.}~\bibnamefont {Kongkhambut}},
  \bibinfo {author} {\bibfnamefont {S.}~\bibnamefont {Rao}}, \bibinfo {author}
  {\bibfnamefont {L.}~\bibnamefont {Mathey}}, \bibinfo {author} {\bibfnamefont
  {H.}~\bibnamefont {Ke\ss{}ler}}, \bibinfo {author} {\bibfnamefont
  {A.}~\bibnamefont {Hemmerich}},\ and\ \bibinfo {author} {\bibfnamefont
  {J.~G.}\ \bibnamefont {Cosme}},\ }\bibfield  {title} {\bibinfo {title}
  {{Condensate Formation in a Dark State of a Driven Atom-Cavity System}},\
  }\href {https://doi.org/10.1103/PhysRevLett.130.163603} {\bibfield  {journal}
  {\bibinfo  {journal} {Phys. Rev. Lett.}\ }\textbf {\bibinfo {volume} {130}},\
  \bibinfo {pages} {163603} (\bibinfo {year} {2023})}\BibitemShut {NoStop}%
\bibitem [{\citenamefont {Valencia-Tortora}\ \emph {et~al.}(2023)\citenamefont
  {Valencia-Tortora}, \citenamefont {Kelly}, \citenamefont {Donner},
  \citenamefont {Morigi}, \citenamefont {Fazio},\ and\ \citenamefont
  {Marino}}]{Valencia2023}%
  \BibitemOpen
  \bibfield  {author} {\bibinfo {author} {\bibfnamefont {R.~J.}\ \bibnamefont
  {Valencia-Tortora}}, \bibinfo {author} {\bibfnamefont {S.~P.}\ \bibnamefont
  {Kelly}}, \bibinfo {author} {\bibfnamefont {T.}~\bibnamefont {Donner}},
  \bibinfo {author} {\bibfnamefont {G.}~\bibnamefont {Morigi}}, \bibinfo
  {author} {\bibfnamefont {R.}~\bibnamefont {Fazio}},\ and\ \bibinfo {author}
  {\bibfnamefont {J.}~\bibnamefont {Marino}},\ }\bibfield  {title} {\bibinfo
  {title} {{Crafting the dynamical structure of synchronization by harnessing
  bosonic multilevel cavity QED}},\ }\href
  {https://doi.org/10.1103/PhysRevResearch.5.023112} {\bibfield  {journal}
  {\bibinfo  {journal} {Phys. Rev. Res.}\ }\textbf {\bibinfo {volume} {5}},\
  \bibinfo {pages} {023112} (\bibinfo {year} {2023})}\BibitemShut {NoStop}%
\bibitem [{\citenamefont {{Wagner}}(1975)}]{Wagner1975}%
  \BibitemOpen
  \bibfield  {author} {\bibinfo {author} {\bibfnamefont {M.}~\bibnamefont
  {{Wagner}}},\ }\bibfield  {title} {\bibinfo {title} {{{A nonlinear
  transformation of SU(3)-spin-operators to bosonic operators}}},\ }\href
  {https://doi.org/10.1016/0375-9601(75)90319-9} {\bibfield  {journal}
  {\bibinfo  {journal} {Physics Letters A}\ }\textbf {\bibinfo {volume} {53}},\
  \bibinfo {pages} {1} (\bibinfo {year} {1975})}\BibitemShut {NoStop}%
\bibitem [{\citenamefont {Emary}\ and\ \citenamefont
  {Brandes}(2003)}]{Emary2003}%
  \BibitemOpen
  \bibfield  {author} {\bibinfo {author} {\bibfnamefont {C.}~\bibnamefont
  {Emary}}\ and\ \bibinfo {author} {\bibfnamefont {T.}~\bibnamefont
  {Brandes}},\ }\bibfield  {title} {\bibinfo {title} {{Chaos and the quantum
  phase transition in the Dicke model}},\ }\href
  {https://doi.org/10.1103/PhysRevE.67.066203} {\bibfield  {journal} {\bibinfo
  {journal} {Phys. Rev. E}\ }\textbf {\bibinfo {volume} {67}},\ \bibinfo
  {pages} {066203} (\bibinfo {year} {2003})}\BibitemShut {NoStop}%
\bibitem [{\citenamefont {Keßler}\ \emph {et~al.}(2020)\citenamefont
  {Keßler}, \citenamefont {Cosme}, \citenamefont {Georges}, \citenamefont
  {Mathey},\ and\ \citenamefont {Hemmerich}}]{Kessler2020}%
  \BibitemOpen
  \bibfield  {author} {\bibinfo {author} {\bibfnamefont {H.}~\bibnamefont
  {Keßler}}, \bibinfo {author} {\bibfnamefont {J.~G.}\ \bibnamefont {Cosme}},
  \bibinfo {author} {\bibfnamefont {C.}~\bibnamefont {Georges}}, \bibinfo
  {author} {\bibfnamefont {L.}~\bibnamefont {Mathey}},\ and\ \bibinfo {author}
  {\bibfnamefont {A.}~\bibnamefont {Hemmerich}},\ }\bibfield  {title} {\bibinfo
  {title} {{From a continuous to a discrete time crystal in a dissipative
  atom-cavity system}},\ }\href {https://doi.org/10.1088/1367-2630/ab9fc0}
  {\bibfield  {journal} {\bibinfo  {journal} {New Journal of Physics}\ }\textbf
  {\bibinfo {volume} {22}},\ \bibinfo {pages} {085002} (\bibinfo {year}
  {2020})}\BibitemShut {NoStop}%
\bibitem [{\citenamefont {Johansen}\ \emph {et~al.}(2023)\citenamefont
  {Johansen}, \citenamefont {Lang},\ and\ \citenamefont
  {Piazza}}]{johansen2023role}%
  \BibitemOpen
  \bibfield  {author} {\bibinfo {author} {\bibfnamefont {C.~H.}\ \bibnamefont
  {Johansen}}, \bibinfo {author} {\bibfnamefont {J.}~\bibnamefont {Lang}},\
  and\ \bibinfo {author} {\bibfnamefont {F.}~\bibnamefont {Piazza}},\
  }\href@noop {} {\bibinfo {title} {The role of atomic interactions in
  cavity-induced continuous time crystals}} (\bibinfo {year} {2023}),\ \Eprint
  {https://arxiv.org/abs/2310.16661} {arXiv:2310.16661 [cond-mat.quant-gas]}
  \BibitemShut {NoStop}%
\bibitem [{\citenamefont {Ke\ss{}ler}\ \emph {et~al.}(2021)\citenamefont
  {Ke\ss{}ler}, \citenamefont {Kongkhambut}, \citenamefont {Georges},
  \citenamefont {Mathey}, \citenamefont {Cosme},\ and\ \citenamefont
  {Hemmerich}}]{Kessler2021}%
  \BibitemOpen
  \bibfield  {author} {\bibinfo {author} {\bibfnamefont {H.}~\bibnamefont
  {Ke\ss{}ler}}, \bibinfo {author} {\bibfnamefont {P.}~\bibnamefont
  {Kongkhambut}}, \bibinfo {author} {\bibfnamefont {C.}~\bibnamefont
  {Georges}}, \bibinfo {author} {\bibfnamefont {L.}~\bibnamefont {Mathey}},
  \bibinfo {author} {\bibfnamefont {J.~G.}\ \bibnamefont {Cosme}},\ and\
  \bibinfo {author} {\bibfnamefont {A.}~\bibnamefont {Hemmerich}},\ }\bibfield
  {title} {\bibinfo {title} {{Observation of a Dissipative Time Crystal}},\
  }\href {https://doi.org/10.1103/PhysRevLett.127.043602} {\bibfield  {journal}
  {\bibinfo  {journal} {Phys. Rev. Lett.}\ }\textbf {\bibinfo {volume} {127}},\
  \bibinfo {pages} {043602} (\bibinfo {year} {2021})}\BibitemShut {NoStop}%
\end{thebibliography}%

\end{document}